\documentclass[a4paper,fleqn,usenatbib]{mnras}
\usepackage[T1]{fontenc}
\usepackage{ae,aecompl}
\usepackage[usenames, dvipsnames]{color}
\usepackage{subfig}

\usepackage{graphicx}	
\usepackage{amsmath}	
\usepackage{amssymb}	

\title[Stellar populations and cluster dynamical stage]{
The dynamic stage of clusters and its influence on the stellar populations of galaxies
}

\author[N. R. Soares \& S. B. Rembold.]{
N. R. Soares,$^{1}$\thanks{E-mail: ruschelsoares@gmail.com.br}
S. B. Rembold$^{1}$
\\
$^{1}$Universidade Federal de Santa Maria - 97105-900, Santa Maria-RS, Brazil\\
}

\date{Accepted XXX. Received YYY; in original form ZZZ}

\pubyear{2017}

\begin{document}
\label{firstpage}
\pagerange{\pageref{firstpage}--\pageref{lastpage}}
\maketitle

\begin{abstract}

We investigate the stellar populations of galaxies in clusters at different dynamical stages, aiming to identify possible effects of the relaxation state of the cluster or subcluster on the star formation histories of its galaxies. We have developed and applied a code for kinematic substructure detection to a sample of 412 galaxy clusters drawn from the \citet[]{tempel2012} catalogue, finding a frequency of substructures of 45\%. We have extracted mean stellar ages with the {\sc{starlight}} spectral synthesis code
applied to SDSS-III spectra of the sample galaxies. We found lower mean stellar ages in unrelaxed clusters relative to relaxed clusters. For unrelaxed clusters, we separated primary and secondary subhalos and found that, while relaxed clusters and primaries present similar masses and age distributions, secondaries present younger stellar populations, mainly due to low-mass galaxies ($\log M_\star/M_\odot \lesssim 11$\,dex). An age-clustercentric radius relation
is seen for all subhalos irrespective of the presence of substructures. We also observe relations between the mean stellar age and mass of relaxed and unrelaxed clusters, massive systems presenting higher mean ages. The locus of these relations is distinct between relaxed and unrelaxed clusters, but become
indistinguishable when separating primaries and secondaries. Our results suggest that differences between relaxed and unrelaxed clusters are mainly driven by low-mass systems in the clusters outskirts, and that, while pre-processing can be seen in the subcomponents of dynamically young clusters, some evolution in the stellar populations must occur during the clusters relaxation.

\end{abstract}




\section{Introduction}

In the current scenario of hierarchical formation, galaxy clusters are formed through a series of mergers of smaller systems and are the most massive structures that  have decoupled from the expansion of the universe \citep[]{press1974,joever1978,vandenbergh1999,kauffmann1999}. In this scenario, clusters of galaxies grow by accretion of both field galaxies and bounded structures like low-mass groups. The process of accretion of lower mass systems by a cluster -- or even the build up of a single massive cluster from merging of two individual clusters -- is known to imprint distinct signatures in the cluster properties before full virialization of the system. These signatures, in the form of asymmetric, extended X-ray distributions \citep{kapferer06,ferrari2006,Parekh+15}, multiple peaks in the galaxy density field \citep[e.g.][]{bbg09,barrena09}, or distinctive line-of-sight galaxy velocity distributions \citep[e.g.][]{ribeiro10,barrena+14}, are usually referred to as ``substructures''. Clusters and groups can be dynamically ``relaxed'' or ``unrelaxed'' according to the presence or absence of substructures. Besides being an important tool for evaluating models of formation and evolution of large scale structures in the Universe \citep{natarajan_wolker04,gao+04,vandenbosch_jiang16,schwinn+17,natarajan+17}, the evolutionary stage of a galaxy cluster is important in understanding the evolution of the baryons (gas and stars) in the galaxies themselves.

It is well known that in the central regions of clusters of galaxies the most common morphology of galaxies found are early-type \citep[]{hogg2003,zhu2010,jaffe2016,deshev2017, sybilska2017} and dominated by older stellar populations \citep[]{bower1999} when comparing to field galaxies, which are mostly late-types with a large contribution of younger stellar populations \citep[]{dressler1980,kelkar2017,oh18}.

This dichotomy can arise both by the initial conditions in which galaxies were first formed and/or by the later transformation of the galaxy morphology and stellar population properties due to the environment. While recent works demonstrate that at least part of the present-time galaxy properties are set early in the cosmic history, favouring the so-called ``nature'' scenario \citep[e.g.][]{Athanassoula10,shi+17}, there is mounting evidence suggesting that the environment is important in the transformation of the galaxy stellar content \citep{poudel+17,hwang18,crone18}. All in all, high density regions are associated to the quenching of the star formation in gas-rich galaxies and subsequent morphological transformations associated to the loss of a gaseous disk. This quenching process may not be limited to the high-density cluster environment, as a large body of observational evidence have shown that infalling galaxies in the clusters outskirts present  properties that are markedly distinct from those of field galaxies, what suggests that such late arrivals have already been pre-processed \citep{wetzel13,hou14}. An extensive zoology of environmentally-driven evolutionary processes have been proposed theoretically to account for the star formation quenching in high-density environments and pre-processing signatures, and  have been confronted with the observations with variable degrees of success. These processes include ram-pressure stripping, tidal stripping, strangulation and galaxy harassment \citep{fujita1999,ruggiero2017,bnm+00,pmc+15,parkhwang09,hvi12}. These effects are thought to present different sensitivities to the environment and occur in different timescales, so a combination of them can be necessary to produce the observed environmental trends \citep[e.g.][]{bahe+13,cora18}.

The environmental density at the location of a given galaxy does not seem, however, to be the only relevant parameter shaping its stellar population properties. It has been found that galaxy properties are distinct between relaxed and unrelaxed clusters, though the details are still unclear. \citet{ribeiro+13a} have shown, for a large sample of low- and high-mass clusters, that relaxed clusters, characterized by a Gaussian velocity distribution, present a higher fraction of red faint galaxies than unrelaxed clusters, while a color-radius relation was detected for both relaxed and unrelaxed structures. In \citet{ribeiro+13b}, on the other hand, no significant radial segregation of galaxy properties have been found for unrelaxed clusters, and most of the differences in the stellar population properties between relaxed and unrelaxed clusters was found to occur for low-mass galaxies; this last result was also found by \citet{carollo+13}, but the intensity of the detected signal was weak. \citet{cohen+14} and \citet{cohen+15} have found a higher star formation rate in galaxies residing in unrelaxed clusters. \citet{decarvalho+17} have found a larger fraction of old, high-metallicity faint galaxies in the outskirts of unrelaxed clusters as compared to relaxed ones, what was interpreted as due to infall of galaxies pre-processed in groups. \citet{robertsparker17}, on the other hand, have found no significant difference in the stellar populations of galaxies outside the virial radius of relaxed and unrelaxed clusters, but at smaller clustercentric distances unrelaxed clusters present a higher star formation rate than relaxed clusters.
\citet{hou2012} have found an increase in the blue population and star forming galaxies in clusters where kinematic substructures, detected by means of the Dressler-Schectman test \citep{dressler1988}, are present. \citet{guennou+14}, identifying substructures in clusters by means of X-ray imaging, have found that substructures which are located near the cluster centers are depleted in late-type galaxies with recent bursts of star formation. 

Most of these works do not directly address if the infall or cluster merging processes are helping to shape the star formation history of the galaxies. A number of works investigating binary pairs of merging clusters \citep[e.g.][]{ferrari2005,hwang2009,stroe+15a,wch+15} have found evidences which suggest an enhancement of star formation in galaxies. These results are often interpreted as the triggering of star formation in an infalling gas-rich galaxy due to shocks in the intracluster medium (ICM) or ram pressure \citep{boc10,ebeling14,roediger14}, though other works suggest that the high merger-driven ICM densities result in quenching of star formation \citep[e.g.][]{fujita+99,boselli06}.
\citet{mansheim+17} have found some hints of recent star formation episodes that could be related to the merging process occurring in DLSCL J0916.2+2953, but this evidence is inconclusive. \citet{ma+10} have found evidence for a merging-induced starburst following the pericentric passage of MACSJ0025.4-1225, resulting in a distribution of E+A galaxies along the merging axis. However, \citet{deshev2017} have found evidence for quenching along the merger axis of the merging cluster A520 but no sign of merging-induced starbursts in the gas-rich galaxy population. A similar result was obtained by \citet{pranger+13} and \citet{pranger+14} for the merging clusters Abell and 3921 Abell 2384 respectively. It is, therefore, still open to debate the impact of the merging process itself in unrelaxed clusters on the stellar population properties, and how much of the observed differences between the stellar population properties of galaxies in relaxed and unrelaxed clusters can be attributed solely to the late arrival of infalling halos. In particular, how are the stellar population in substructures compared to relaxed clusters of similar mass? Do individual substructures in unrelaxed clusters present evidence of pre-processing? What is the relative impact in relaxed and unrelaxed clusters on the SFH of galaxies with different stellar masses? Does the pre-processing occur in all halo mass scales?
 
In this work, we investigate differences between stellar populations in galaxies in clusters with different dynamic stages, aiming to identify the influence of environmental effects in the galaxy stellar populations by the decomposition of unrelaxed clusters in their constituent subhalos, trying to shed some light on the above questions. For this purpose, we use a sample of clusters of groups and clusters selected from the \citet[]{tempel2012} catalogue, based on the  eighth data release of SDSS III. To identify substructures and characterize their kinematic properties, masses and radii, we developed an automated algorithm based on the $k$-test of \citet[]{Colles1996}. We use the {\sc{starlight code}} \citep[]{cid2005,mateus2006} to perform the stellar population synthesis and derive the mean stellar age of galaxies.

The paper is organized as follows: in Section 2 we describe our data and catalogue; in Section 3 we describe our methodology for substructure identification and characterization, and also the derivation of the stellar population properties of the galaxies in these structures. In Section 4 we present the results, where we investigate how the stellar populations of galaxies depend on the properties of the structure they reside in, and in Section 5 we discuss and summarize our conclusions. In this work we assumed a cosmology with $\Omega_m$=0.3, $\Omega_{\Lambda}$=0.7 and H$_{0}$= 70 kms$^{-1}$ Mpc$^{-1}$

\section{Data}
{\subsection{Cluster and group sample}}
In this work, we intend to analyze the stellar populations of galaxies in groups
and clusters spanning the largest range possible in richness and dynamic stage. For our
purposes, it is convenient to rely on a large, homogeneous catalogue, with a strict
identification of confirmed members and interlopers (i.e. free from projection effects),
and for which optical spectra are available. The \citet[]{tempel2012} catalogue (hereafter TTL) is a large spectrophotometric catalogue implemented on the eighth data release of Sloan Digital Sky Survey \citep{aihara+11}. This catalogue contains 77,858 groups with more than three confirmed members and spanning the range $0.009<z<0.2$. The groups are identified in the redshift-projected position space with the friends-of-friends algorithm with a variable linking length to avoid selection effects. As we describe in Sect. 3, the characterization of the dynamical stage of a cluster using galaxy velocities relies on a large number of spectroscopically confirmed members. We have thus selected from this catalogue all clusters with more than 30 galaxies, what resulted in a sample of 412 clusters, comprising 24,169 galaxies. From this preliminary list, we excluded eight objects (identified by the codes 09349, 13462, 20575, 24918, 33262, 34727, 49298 and 62138) which correspond to superclusters and filaments and are too complex for our analysis.

Using the CASJOBS server\footnote{http://skyserver.sdss.org/casjobs/}, we have obtained the SDSS-III optical spectra for all galaxies in our sample from Data Release 10 \citep{ahn+14}. These single-fiber spectra cover the wavelength region 3800-9200 {\AA} with a spectral resolution $\sim$2000 and cover the inner 3 arcseconds of the galaxy. We have also obtained the absolute magnitudes in the $r$ band as derived from the SDSS photometric redshift pipeline. Because some galaxies lack an estimate of photometric redshift, the final sample of cluster galaxies was slightly reduced to 20,192 galaxies.

\begin{figure*}
\centering
\includegraphics[width=\columnwidth]{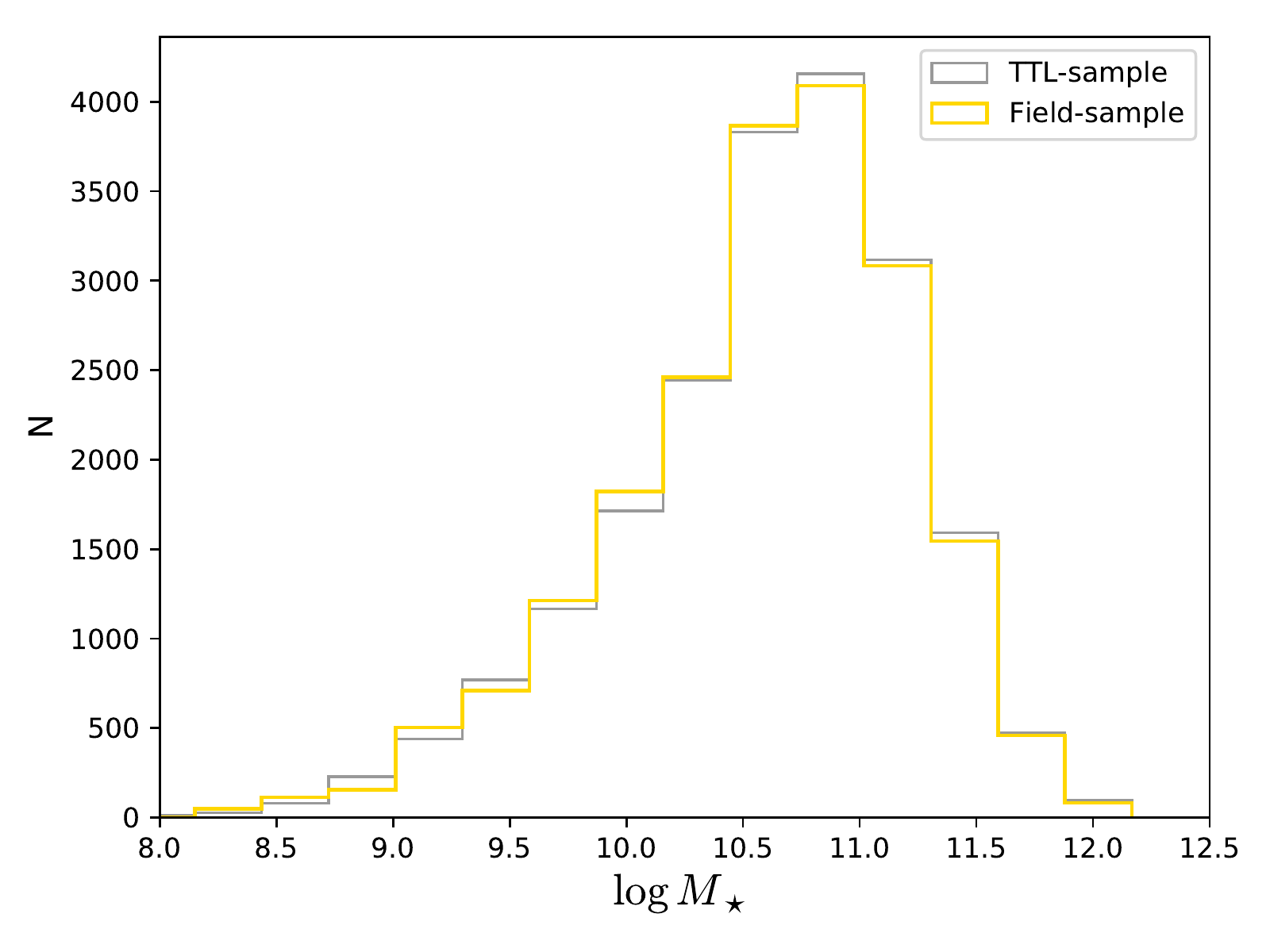}
\includegraphics[width=\columnwidth]{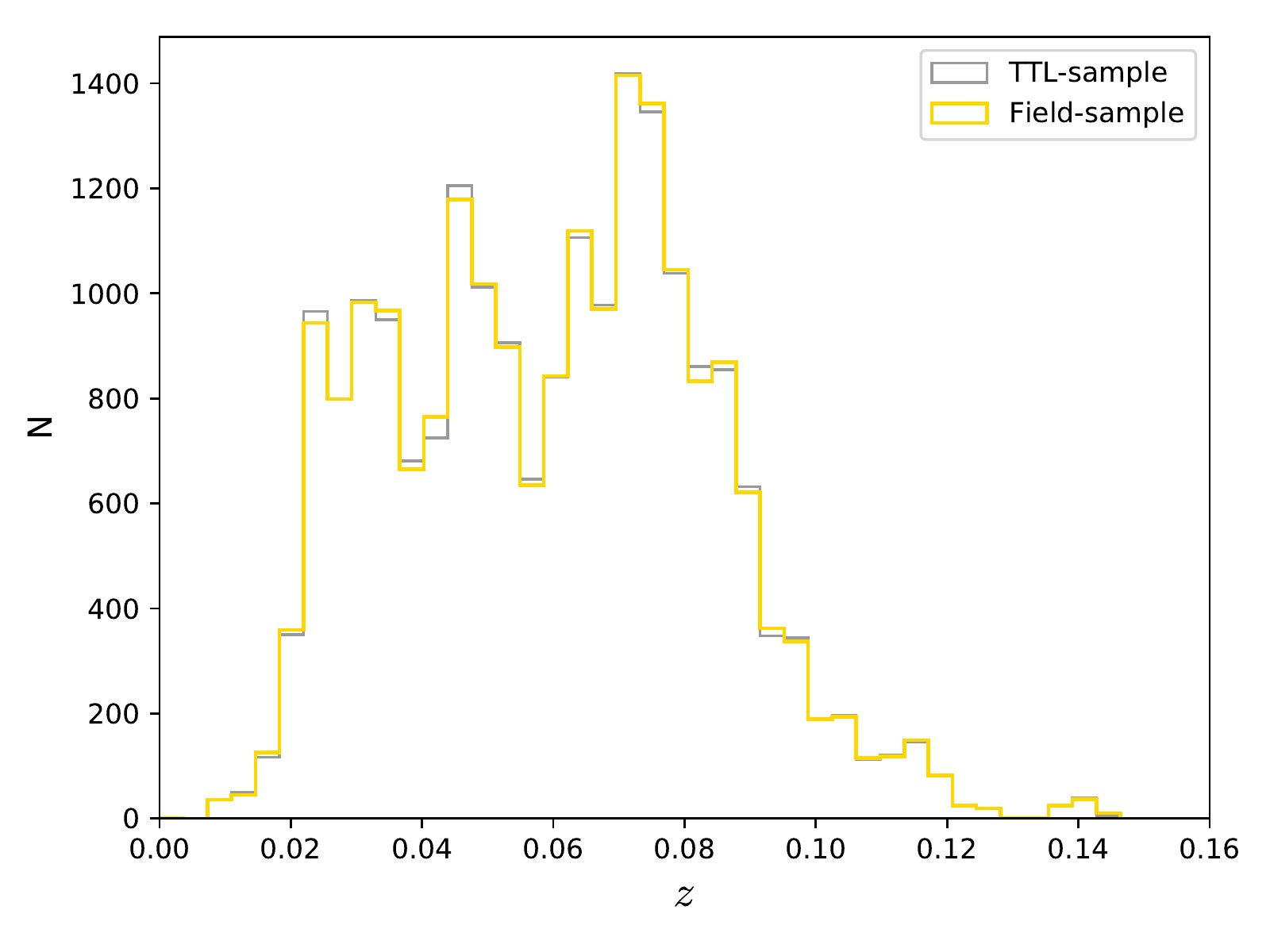}
\caption{Stellar mass (left) and redshift (right) distributions of galaxies in our cluster (grey) and field (yellow) samples.}
\label{mass-z}
\end{figure*}

{\subsection{Field sample}}

In order to better constrain the impact of the cluster evolutionary stage on member galaxies, it is important to compare the general properties of galaxies in clusters with field galaxies. We have therefore selected a control sample of field galaxies -- i.e. galaxies not associated to clusters or groups -- which are otherwise similar to out sample of cluster galaxies in stellar mass and redshift. Using the CASJOBS server, we have selected a preliminary field sample with 200,000 galaxies with the same $r$-band apparent magnitude as the TTL sample and selected from SDSS-III DR-10. We then measured the maximum projected clustercentric distance $D_{max}$ and the maximum peculiar velocity $V_{max}$ of all galaxies in the TTL sample with relation to its parent cluster/group. A galaxy from our preliminary list was then confirmed as a field galaxy if (a) its projected distance to each cluster center exceeds 2$\times D_{max}$ or (b) its line-of-sight velocity offset relative to all clusters in the TTL sample exceeds $V_{max}$+2000 km\,s$^{-1}$. After performing these cuts, the field sample was reduced to 35,402 galaxies. For this preliminary sample, we obtained the SDSS-III single-fiber spectra and the $r$-band values of the absolute, apparent and fiber magnitude. These galaxy spectra were then used to estimate their stellar masses and mean stellar ages (see Sect.~\ref{sect:sps}).

The next step was to compare the stellar mass and redshift of galaxies in field sample with those of galaxies in our cluster sample. For this, we introduce the metric
\begin{equation}
\mu = \sqrt{\left(\frac{z-z_{fld}}{0.05}\right)^{2} + \left(\frac{M_{\star} - M_{\star fld}}{2\times 10^{11.5}}\right)^{2}},
\label{metric}
\end{equation}

\noindent \noindent where $z$ and $M_{\star}$ are the redshift and the stellar mass of galaxies in the cluster sample (sect.~\ref{sect:sps}), and $z_{fld}$ and $M_{\star fld}$ are these same parameters for field galaxies. For every galaxy in the cluster sample we have then selected the field galaxy which minimized $\mu$. The final field sample comprises therefore the same number of galaxies as our sample of cluster galaxies. In Figure~\ref{mass-z} we compare the stellar mass and redshift distributions of galaxies in the field and cluster samples.

\section{Substructure detection}

Our main objective is to analyze the differences between galaxies that reside in clusters in different dynamical stages through their stellar populations, and for this we developed a method that is able to (automatically) detect substructures in galaxy clusters and derive their
kinematic parameters. The Dressler-Schectman ($\delta$) test \citep[]{dressler1988} and the $\kappa$-test \citep[]{Colles1996} are 3D tests that have been used to detect substructures in clusters. However, the identification of each substructure and of the galaxies therein usually relies on the visual analysis of ``bubble plots'', where galaxies are represented as circles proportional to the size of a statistical parameter which is sensitive to differences between the local and the global velocity distributions. In the $\delta$ test, for each galaxy \emph{i} and its $N_n$ nearest neighbors in the projected position space, the $\delta_{i}$ statistics for each \emph{i} galaxy is defined by:

\begin{equation}
\delta_{i}^{2} = \frac {N_n}{\sigma_{g}^{2}} \left[(v_{l}-v_{g})^{2} + (\sigma_{l}-\sigma_{g})^{2}\right]
\label{delta}
\end{equation}

\noindent where $v_l$ ($v_g$) is the local (global) mean velocity, and $\sigma_l$ ($\sigma_g$) is the local (global) velocity dispersion. A local ``cloud'' of large $\delta_i$ values is interpreted as evidence for the presence of a substructure. The summation of all $\delta_i$ values produces the $\Delta$ statistics that, when compared to the typical values obtained from Monte Carlo shuffling the velocities of the galaxies in the cluster, can be used to infer the overall evolutionary stage of the cluster.

The \emph{k}-test of \citet[]{Colles1996} is similar to the $\delta$ test, but the $\kappa$ statistic is derived from the comparison of the full local/global velocity distributions,
as opposed to the $\delta$ statistics and its dependence only on the first two moments
of the distribution. For each $i$ galaxy, the $\kappa_i$ statistics calculated by:

\begin{equation}
k_{i} = -\log P_{KS} (D > D_{obs})
\label{ki}
\end{equation}

\noindent where P$_{KS}$ is the Kolmogorov-Smirnov probability that the local velocity distribution and the global velocity distribution are derived from the same original distribution. The derived ``bubble plots'' and the summation of the $\emph{k}_i$ values are interpreted  as in the delta test. However, the $\kappa$-test has the severe limitation of being non-automated for identifying galaxies in substructures. We have therefore developed an automated algorithm -- \emph{Local Kinematic Estimator} ({\sc{LocKE}}) inspired on the $\kappa$-test that is capable of identify the individual structures and assign the galaxies to the proper structure, without visual inspection.

\subsection{LocKE - Local Kinematic Estimator}

\begin{figure*}
\centering
\subfloat[]{\label{f223-10}\includegraphics[width=\columnwidth]{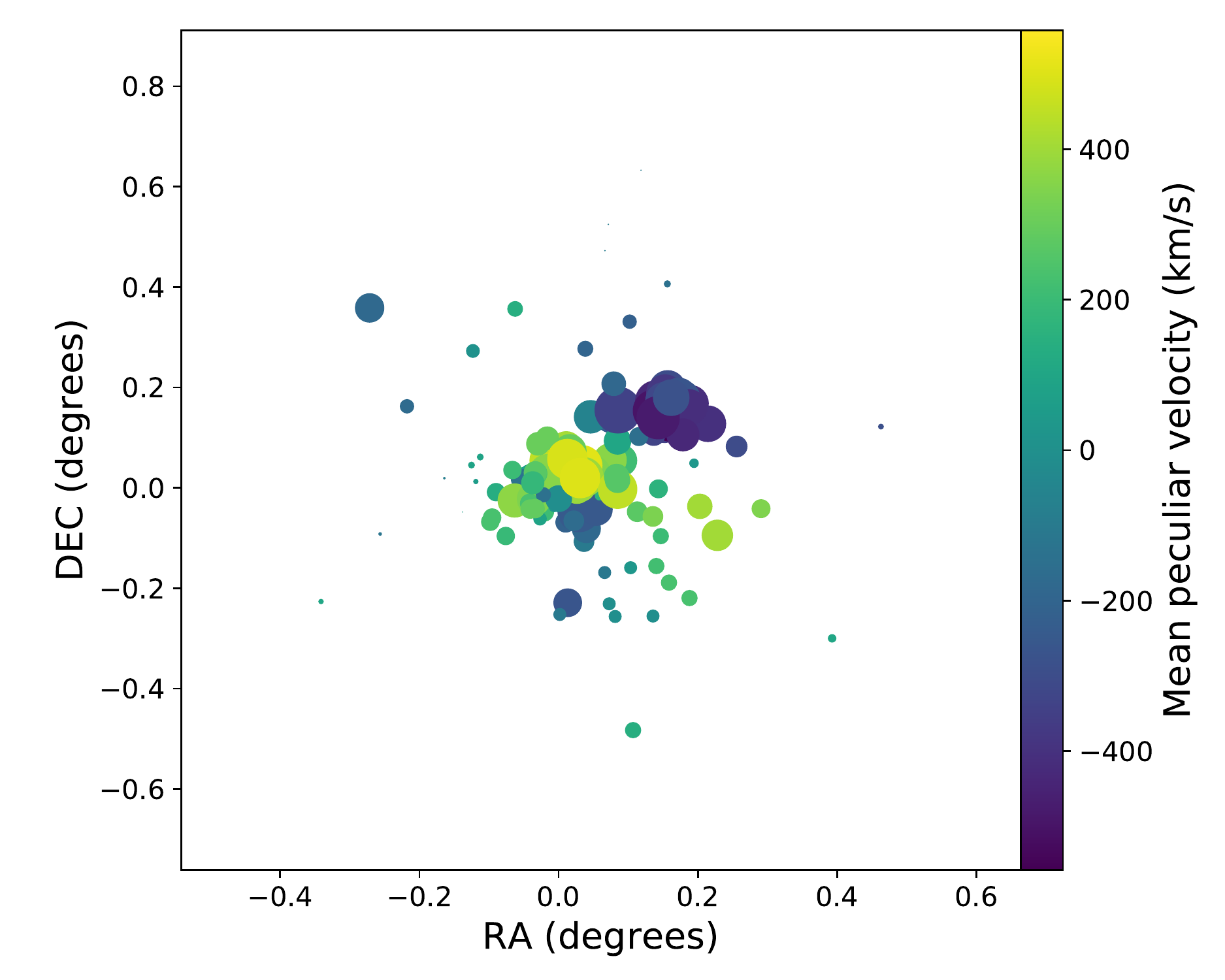}}
\subfloat[]{\label{f223-30}\includegraphics[width=\columnwidth]{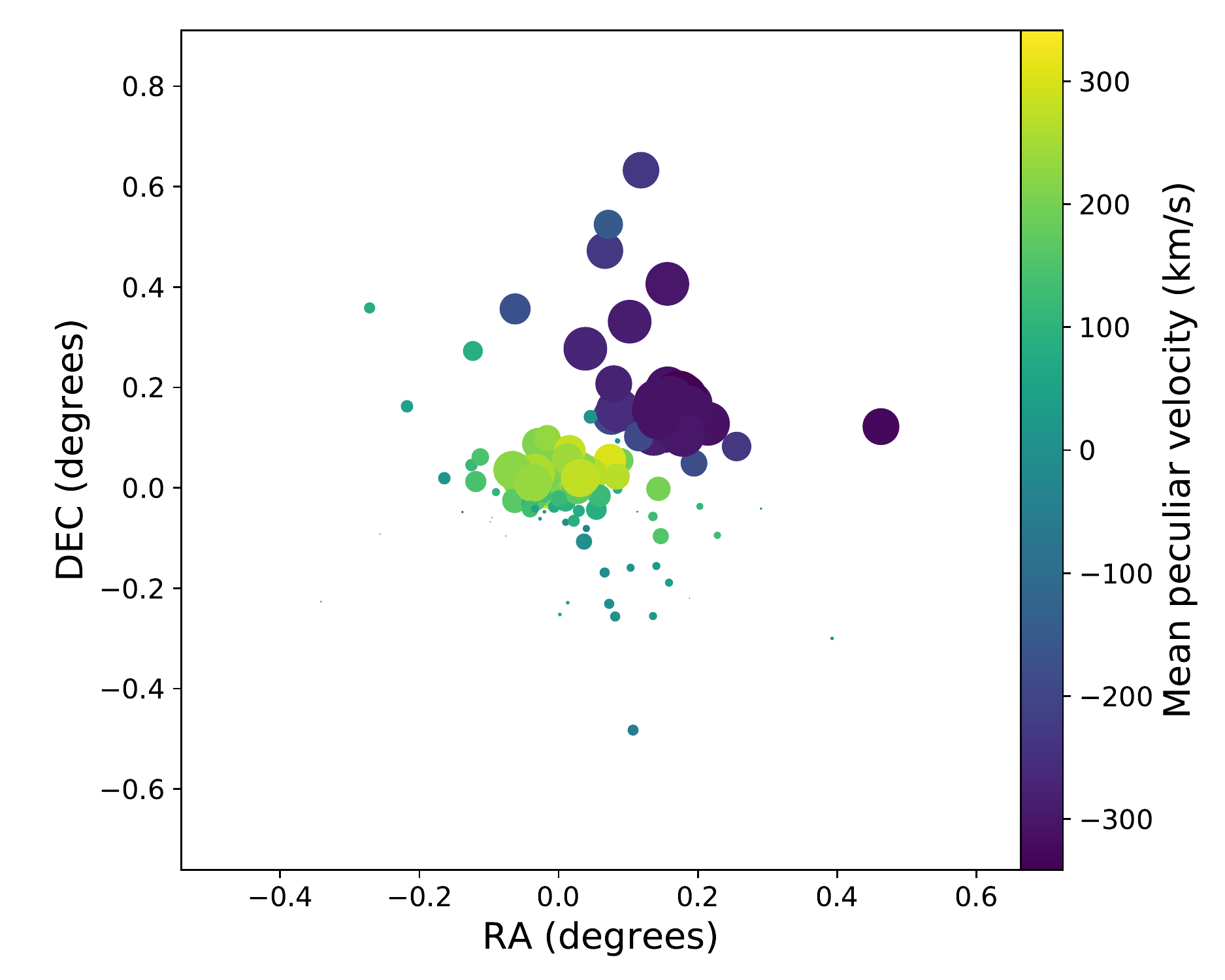}}\\
\subfloat[]{\label{f223-50}\includegraphics[width=\columnwidth]{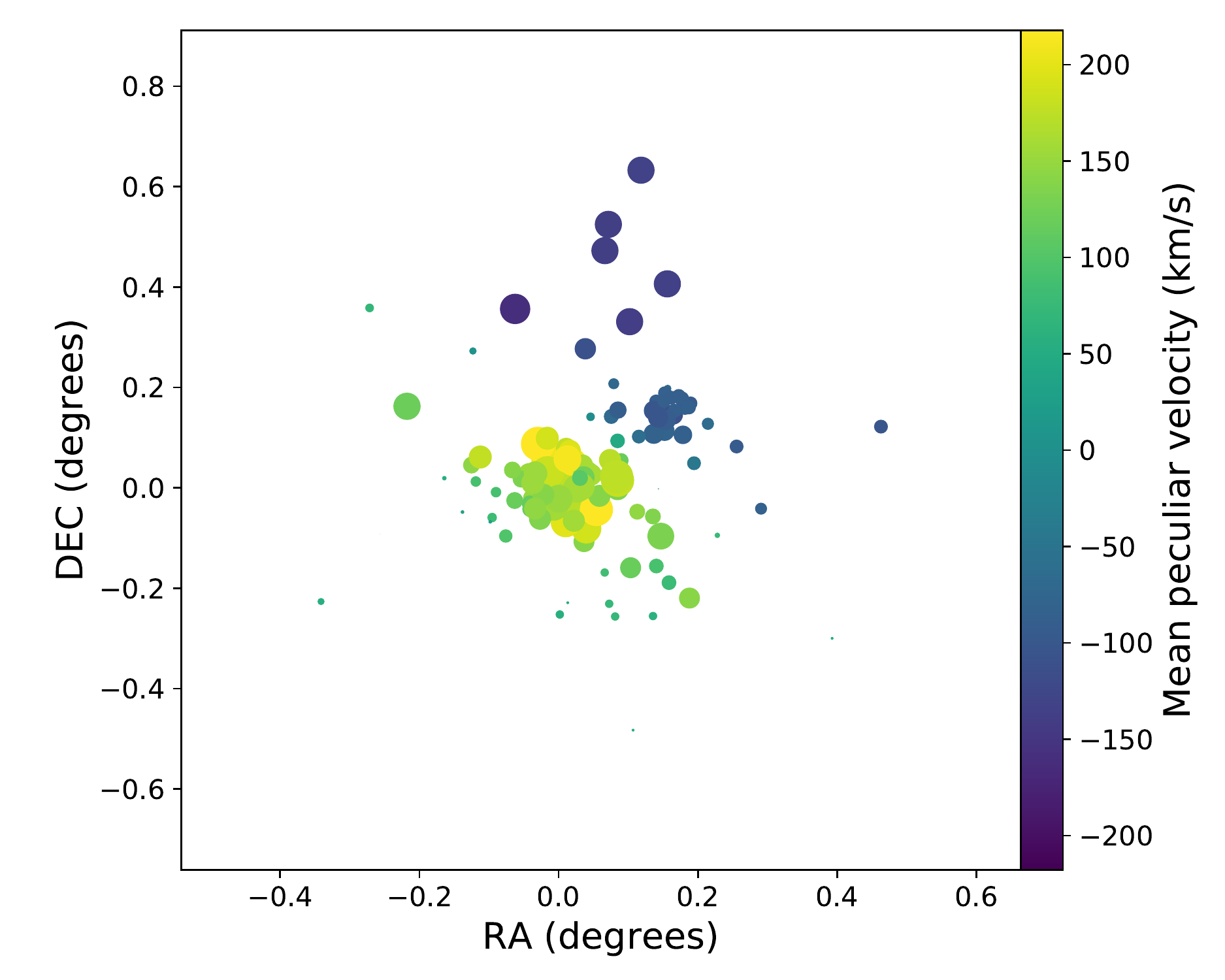}}
\subfloat[]{\label{f223-70}\includegraphics[width=\columnwidth]{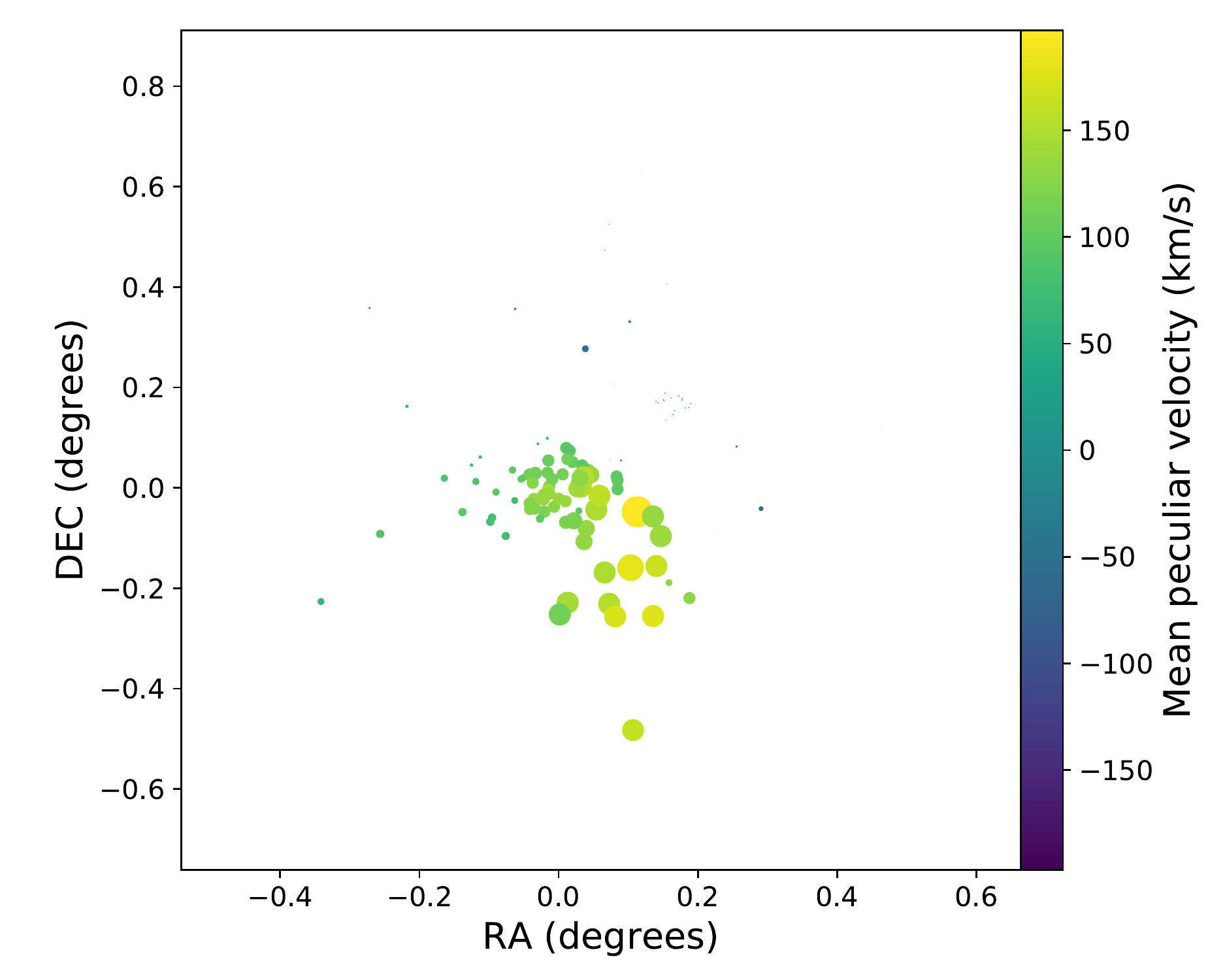}}
\caption{Bubble plots for a bimodal simulated cluster, for $N_n=10$ (a), $30$ (b), $50$ (c) and $70$ (d). The size of each circle is related to the size of the $\kappa$
statistics, and the circle colours indicates the average line-of-sight velocity of the $N_n$ neighbours.}
\label{f223}
\end{figure*}

The main goal of {\sc{LocKE}} is to extract individual substructures and their member galaxies, providing estimates of their kinematic parameters, through comparison between the velocity distributions in different regions of the clusters. The success of this procedure is crucially dependent on the contrast between the local and global velocity distributions, and therefore on the definition of ``local'' and ``global'' galaxies -- i.e. the choice
of the number of neighbours $N_n$. \citet[]{dressler1988} used a fixed value of $N_n$ at 10, while \citet[]{Aguerri2010} used the square root of the total number of galaxies in the cluster. This choice is crucial for the correct detection of substructures and,
as a general rule, the largest statistics in the $\kappa$ test, and the largest contrast
between the substructure and the remainder of the cluster in the ``bubble plots'' will be obtained when $N_n$ is comparable to the number of galaxies in the substructure. This is illustrated in Figure~\ref{f223}, where we present the $\kappa$-test ``bubble plot'' for
a synthetic bimodal cluster composed of a primary structure of 92 galaxies with $\sigma=878$\,km/s and a secondary structure of 32 galaxies with $\sigma=518$\,km/s and
a line-of-sight velocity offset of $563$\,km/s. We have run the $\kappa$ test with
variable $N_n$ values. For each run we show in the plots the summation of the $\kappa$ values and its corresponding percentile after 1000 realizations of Monte Carlo shuffling in the velocity space -- i.e. the significance of substructure detection. Notice that both the $\kappa$ statistics and the significance of the test are lower, and the location and physical limits of the substructure become harder to define, for $N_n$ much larger or smaller than the number of galaxies in the substructure.

\begin{figure*}
\centering
\subfloat[]{\label{f345-30a}\includegraphics[width=\columnwidth]{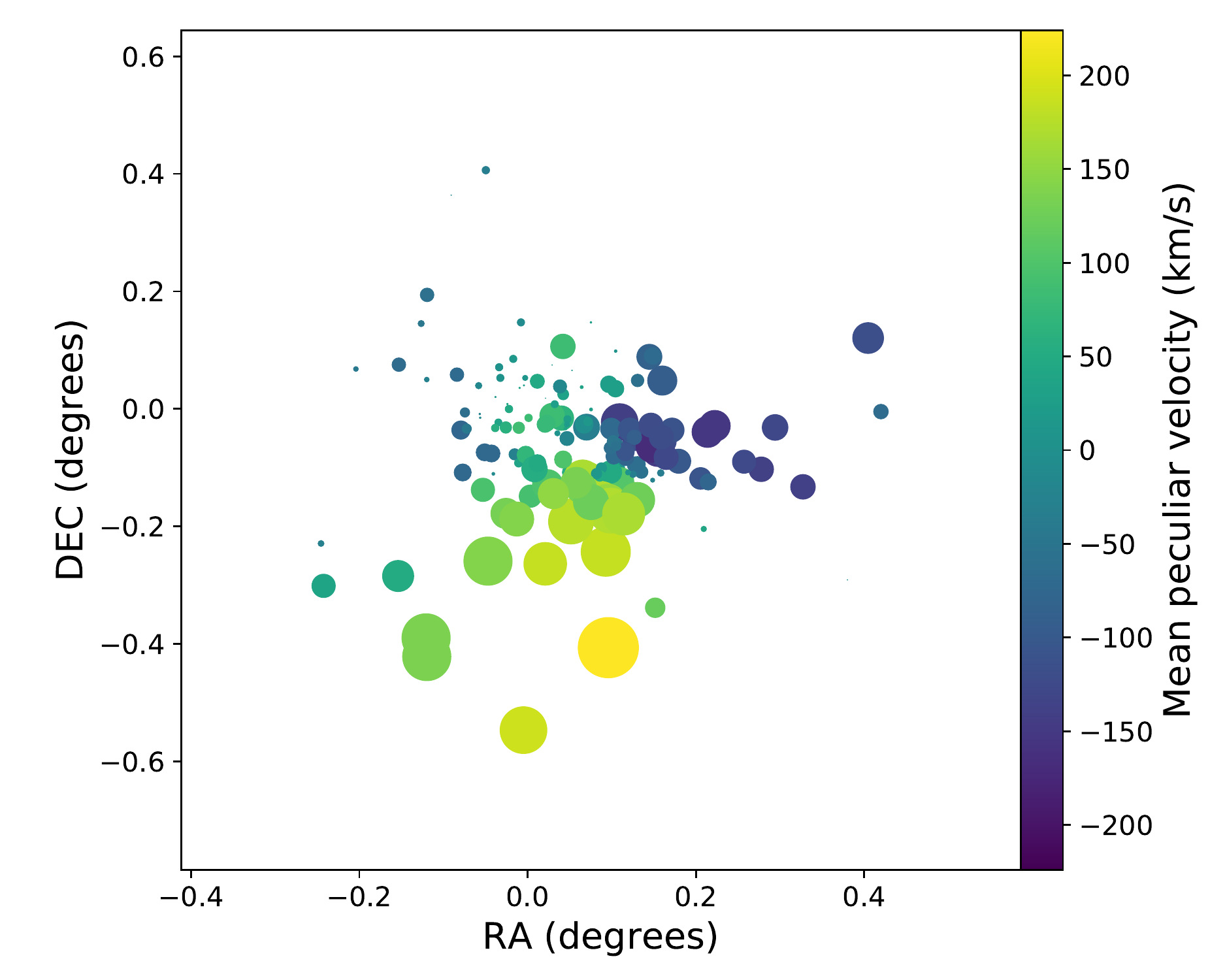}}
\subfloat[]{\label{f345-30b}\includegraphics[width=\columnwidth]{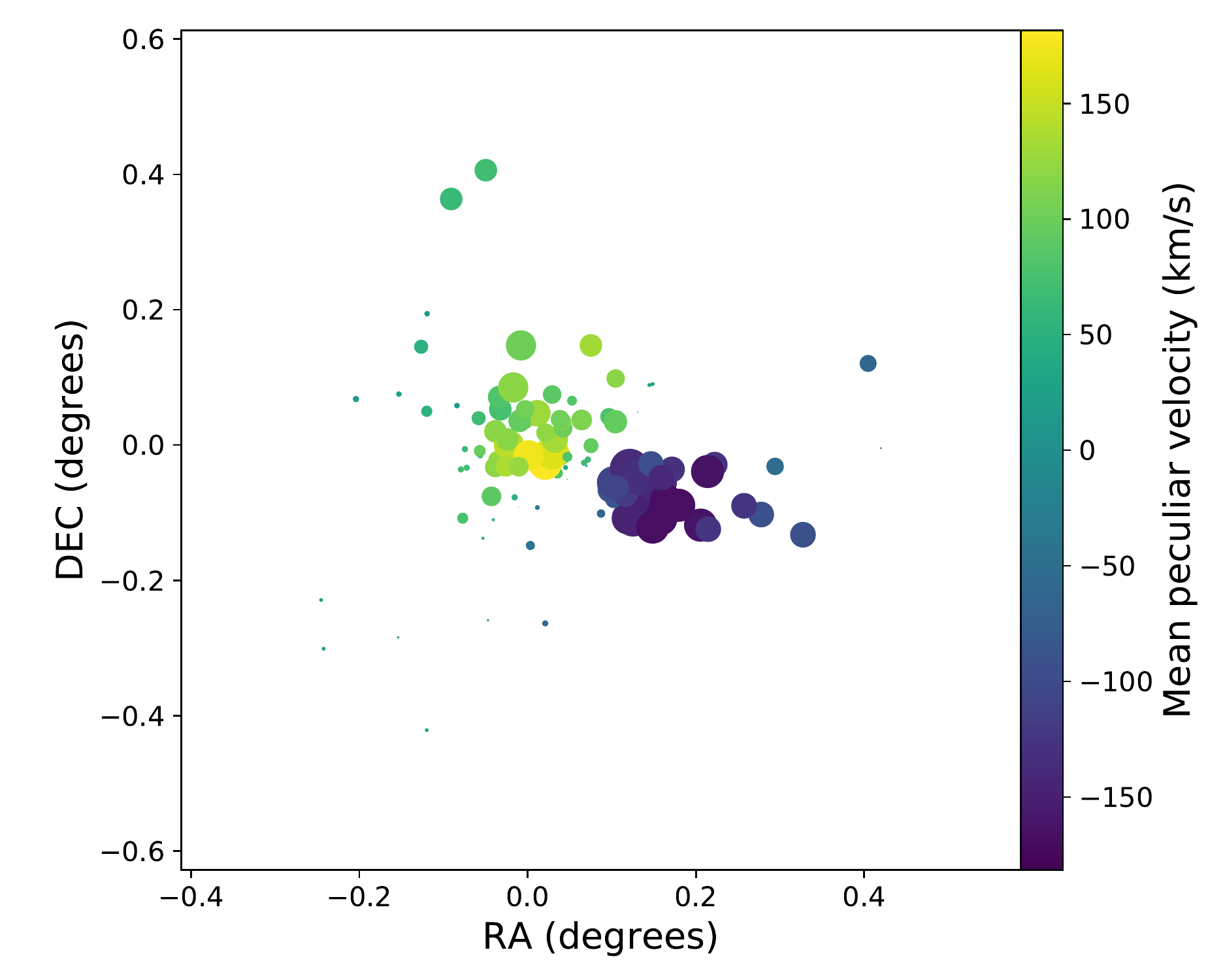}}
\caption{Bubble plots for a simulated cluster with three kinematic structures. In panel (a),
all galaxies are included. In panel (b), we exclude the structure associated to the larger $\kappa$ values.}
\label{f345}
\end{figure*}

The above discussion does not take into account the presence of multiple substructures in a cluster. For a multimodal cluster comprised of one main structure and a number of secondary structures, the largest $\kappa$ values in general will be obtained for the substructure with the larger kinematic differences relative to the full velocity distribution, but other structures, when present, can usually also be visible at smaller or comparable $\kappa$ values. Individual
substructures must therefore me identified in a case-by-case analysis which take into consideration
the typical $\kappa$ values of each region and multiple concentrations of galaxies. We illustrate this
in Figure~\ref{f345}, which presents the bubble plot of a simulated cluster made up of three individual kinematic structures. Notice that the two clouds of large and medium-sized $\kappa$ values are only identifiable as two structures by comparison of the local mean velocities, but not by the size of the $\kappa$ statistics. When we exclude the region associated to the largest $\kappa$
values, the bimodal structure of the remaining field becomes obvious.

The examples above show that, in order to automatically identify individual substructures in the field of a cluster
using the $\kappa$ test, a stratified, multi-scale comparative analysis between the local and global kinematics
of a cluster is mandatory. That is the rationale behind {\sc{LocKE}}, developed in the Python language and whose steps and structure we describe in the following.

\subsubsection{Initialization}

{\sc{LocKE}} expects as input a list of sky coordinates and redshifts of the $N_o$ cluster member galaxies. At initialization, the nearest neighbours of each galaxy are obtained using the \texttt{BallTree} algorithm implemented in the Python module \texttt{scikit-learn} (Pedregosa et al. 2011).

\subsubsection{Statistics maximization}

We start with a first value for the number of neighbours $N_n=\sqrt{N_o}$, which we define as the lowest possible number of neighbours which define a substructure in the $\kappa$-test bubble plots. {\sc{LocKE}} then creates internally a grid of 10 equally-spaced values of $N_n$, ranging from $\sqrt{N_o}$ to $3N_o/4$. For each of these values, the individual $\kappa_i$ values are obtained for all cluster galaxies. For this calculation, we define as the ``global'' kinematics the velocity distribution of all galaxies \emph{except} the $N_n$ nearest neighbours of each galaxy, instead of the full cluster velocity distribution. This is done to maximize the absolute value for the $\kappa$ statistics. {\sc{LocKE}} then identifies the number of neighbours $N_{n,max}$ for which the average value of $\kappa$ is maximum. The spacing between consecutive values of the $N_n$ grid is then reduced to one half of its initial value, and the $\kappa_i$ calculation is repeated, producing an updated value of $N_{n,max}$. This procedure of reducing the spacing of the $N_n$ grid is repeated until convergence. The convervence value of $N_{n,max}$ therefore corresponds to the number of neighbours which produces the best contrast between local and global velocity distribution. We will refer to the convergence value of $N_{max}$ as $N_{n,best}$. Notice that, at this stage, we allow a ``substructure candidate'' to be composed of more than half of the total number of galaxies in the cluster field; we will later discuss the reason behind this choice.

\subsubsection{$\kappa$ statistics}

For the $N_{n,best}$ nearest neighbours, {\sc{LocKE}} performs a final calculation of the
$\kappa_i$ values, as well as the summation of these values, $\kappa_{clus}$. {\sc{LocKE}} then randomly shuffles the velocities of all galaxies in the cluster field and, using the same $N_{n,best}$ value, calculates the respective $\kappa$ values. The significance of the observed $\kappa_{clus}$ statistics is obtained by comparison with the distribution of $\kappa$ values obtained for a large number of reshuffling runs; we consider a positive detection of substructure all cases where the significance of the test is higher than 95\%. In our early experiments we have found that, for a large number of clusters, the significance of the $\kappa$ test is barely higher or lower than 95\%, so that a very large ($\sim 10^5$) number of simulations had to be performed to confirm a discard the presence of substructures in these clusters. So, instead of performing so many simulations for all clusters, even those which are clearly devoid of substructures, we follow a different approach. For $N_s$ blocks of 20 simulations each, we calculate the average 95\% percentile $p$ of all simulated $\kappa_i$ values, and the standard error $\sigma$ of this percentile. We then increase $N_s$ until the absolute difference between $p$ and the observed $\kappa_{clus}$ value is larger than $2\sigma$ -- i.e. until $p$ itself is constrained at the 95\% level. For clusters with evident substructures or lack thereof, $N_s$ has shown to be as low as $\sim 3-4$, resulting in a very short processing time; for other clusters, however, the $N_s$ value can be of the order $\sim 10^2$, and the processing time will be correspondingly higher. In our tests, the identification (or not) of substructures using this criterion was rigorously identical as that obtained by simply assuming a fixed large number of reshuffling operations. It is important to note that all that {\sc{LocKE}} cares about is whether or not a substructure is detected in the field of a cluster, independently of the particular value of its statistical significance. 

\subsubsection{Absence of substructures}

If the significance of substructuring is lower than 95\%, {\sc{LocKE}} assumes that the
cluster is devoided of substructures. Using the kinematics and positions
of all galaxies in the cluster, it estimates the mean line-of-sight velocity $v$ and dispersion $\sigma_v$,
using the biweight location and scale estimator implemented in the \texttt{astLib.astStats} Python module, assuming tuning constants for location and scale of 6.0 and 9.0, respectively \citep[]{beers1990} . This calculation is done with a 3-$\sigma$ clipping algorithm which is a slight modification of the \texttt{astStats.biweightClipped} module. The uncertainties in both kinematical parameters are obtained with the jackknife technique \citep[]{beers1990}.

\subsubsection{Identification of substructures}

If the significance of substructuring is higher than 95\%, {\sc{LocKE}} assumes that a potential substructure has been found. The spatial distribution of the $\kappa_i$ values is then used to estimate the center and the extension of the substructure detected. The map of galaxy coordinates is convolved with a Gaussian distribution, the standard deviation of which is calculated by the average distance of the $N_{n,best}$ nearest neighbours of all galaxies in the cluster. At this step, the statistical weights of the convolution are the $\kappa_i$ values of each galaxy. This results in a continuous density map where the local density is directly associated to the size of the $\kappa$ statistics around each point. {\sc{LocKE}} then attributes the first-guess center of the substructure to the peak of the $\kappa$ density map. The first-guess substructure member ``candidates'', $P$, are all the $N_{n,best}$ nearest neighbours of the galaxy closer to the density peak. If, for any of these $P$ preliminary members, the average $\kappa$ values of its $N_{n,best}$ nearest neighbours is higher than for the galaxy closer to the density peak, the list of candidates $P$ is updated by replacing the galaxy closer to the density peak for the galaxy with the higher $\kappa$ value, and the substructure member candidates are replaced by its $N_{n,best}$ nearest neighbours.

\subsubsection{Substructure decontamination}

The member candidates of the substructure, as identified in the previous step, usually include galaxies
which are members of other structures in the field of the cluster. The substructure contamination
will be the more severe the larger the superposition of the individual 3D structures along the line of sight.
{\sc{LocKE}} tries to clean the substructures by means of a two-gaussian mixture modelling in the velocity space,
using the \texttt{mixture} module. The starting parameters of gaussian mixture are the median and
standard deviations of the velocity distributions of the substructure candidates and the remaining galaxies in the cluster.
After convergence, galaxies are ``confirmed'' as substructure members if they are associated to the substructure
velocity distribution by the \texttt{mixture} module; the remaining galaxies -- usually very few, but in some cases a significant fraction of the full substructure -- are considered
outliers and are attributed to the remaining cluster structure. The field contamination is the reason why we allow a potential substructure to contain more than half the total number of galaxies in the field. The parameters of the velocity distribution of the substructure are assumed to be the center and width of the gaussian fit obtained by the \texttt{mixture} module. Uncertainties in these parameters are derived by bootstrapping the velocity distribution of the substructure candidates and running \texttt{mixture} module for all realizations, where the number of bootstrap operations is set by the user.

\subsubsection{Multiple substructures and dynamical parameters}

After identification of the first-level substructure, all steps are repeated in the remaining galaxy field, after excluding the confirmed substructure members. Multiple levels of substructuring can therefore be detected for the same cluster. These steps are repeated until no further substructures are detected, i.e. the significance of the $\kappa$ test is lower than 95\%. Having identified and extracted all individual substructures in the field of a cluster, {\sc{LocKE}} then derives masses and physical radii for each individual structure.

The virial mass ($M_v$) of the structures was measured using equation \ref{massav}, assuming that the galaxy distribution in the structure follows the mass distribution and the system is linked by the same gravitational potential well \citep[]{merritt1988,girardi1998}.

\begin{equation}
M_v= \frac{3\pi}{2} \frac{\sigma_{P}^{2} \ R_{PV}}{G},
\label{massav}
\end{equation}

\noindent here $\sigma_{p}$ is the projected velocity dispersion measured by {\sc{LocKE}} and R$_{PV}$ is the projected virial radius of the structure, given by 

\begin{equation}
R_{PV}= \frac {N(N-1)}{\sum_{i \neq j} R_{ij}^{-1}},
\label{raioproj}
\end{equation}

\noindent \citep[]{girardi1998}, where $R_{ij}$ is the distance between a pair of galaxies $i$ and $j$, and $N$ is the total number of galaxies in the structure. A pressure correction term $C$ was applied on the mass estimate; the lack of this correction will overestimate the cluster masses \citep[]{carlberg1996}. In this work, we measured the corrected virial mass inside a virial radius $R_{vir}$ given by

\begin{equation}
R_{vir} = \sqrt[3]{\frac{\sigma_{P}^{2} R_{PV}}{6 \pi H_0^2}},
\label{virrad}
\end{equation}

\noindent and the corrected mass is

\begin{eqnarray}
\nonumber
M_{CV} & = & M_V - C\\
\nonumber
& = & M_V\left[1-4\pi R_{vir}^{3} \frac{\rho(R_{vir})}{\int^{R_{vir}}_{0}4\pi r^2 \rho dr} \left(\frac{\sigma_r(R_{vir})}{\sigma(<R_{vir})}\right)^2\right],
\label{massacor}
\end{eqnarray}

\noindent where the parameter $\left(\sigma_r(b)/\sigma(<b)\right)^2$ is the velocity anisotropy. The anisotropy parameter is not easy to extract for a given galaxy structure, especially for very poor systems, so that instead of deriving it for each detected structure, we simply assume that all individual structures are dynamically relaxed  -- even though the systems they are part of can be themselves unrelaxed structures. As shown by \citet[]{costa18}, relaxed clusters typically present radially decreasing velocity dispersion profiles. Velocity dispersion profiles of this type are produced by isotropic velocities in the inner region and radial velocities in the external regions, and can be well described by a velocity anisotropy parameter of 0.6 \citep[]{girardi1998}. We have therefore
fixed the anisotropy parameter to this value. 

Finally, the $R_{200}$ radius is calculated following \citet[]{yan2015},

\begin{equation}
R_{200}=\frac{\sqrt{3} \sigma}{10H(z_c)},
\label{R200}
\end{equation}

\noindent where $\sigma$ is the velocity dispersion of the structure and $H(z_c)$ is the Hubble factor at redshift $z_c$.

\subsection{LocKE performance}

A variety of methods for detecting substructures in clusters is available in the literature, but most of these
do not automatically separate galaxies in the many individual cluster structures. One test that can be used for this is multidimensional normal mixture modelling. In this test, the full 3D space of the cluster galaxies is decomposed with a mixture of multidimensional Gaussians. The number of individual Gaussians found in the solution is interpreted as the amount of individual structures detected in the cluster. This multidimensional modelling, when compared to the $\delta$ or $\kappa$ tests, has the advantage that the separation of galaxies in each structure is done automatically. This method has been successfully applied by \citet[]{Einasto2012} to identify structures in a large sample of groups and clusters. The authors have performed the mixture modelling with the package \texttt{Mclust} \citep[]{raftery1999}. We performed some preliminary tests using simulated clusters to determine the performance of \texttt{Mclust} in comparison to the $\kappa$-test. In the performed tests, the $\kappa$-test performed better than \texttt{Mclust} for correctly identifying substructures whose centroids do not overlap spatially; this, in fact, was one of the motivations behind {\sc{LocKE}} development. In the following, we quantify the {\sc{LocKE}} performance taking the \texttt{Mclust} performance as a reference.

For this, we have created a set of 3000 simulated galaxy clusters, containing between 30 and 250 members. The simulations have been created with a maximum of 3 individual structures in a cluster. The simulated clusters are arranged in such a way that 20\% of the sample are clusters with only 1 structure (without substructure), 40\% of the clusters have two structures (a primary and a secondary component) and the remaining 40\% of the clusters sample have 3 structures, a primary component, a secondary component and a tertiary component. Thus, 80\% of the simulated clusters show more than one structure. The number of galaxies in each structure is randomized, imposing the condition that the secondary structure is poorer than the primary and richer than the tertiary. For simplicity, the spatial distribution of galaxies in each structure follows a circularly symmetric radial density profile given by $N(R)=N_0/(1+(R/R_C^2))$ \citep[an analytic approximation to the][profile]{king1962}. The core radius and velocity dispersion of each structure are scaled to its number of members. So that there is no overlap between the structures centroids in the position space, the centroid offsets between the primary and every other component are obtained through a uniform random distribution within $(R_{C1},4R_{C1})$; we do not allow for completely aligned structures along the line of sight because LocKE, just like the usual $\kappa$-test and $\delta$-test, only detects substructures when the velocity distribution and the galaxy projected positions are simultaneously distinct from those of the remaining galaxies in the cluster. The relative velocities between the primary and other structures are randomized and limited to 900\,km/s. The velocity distribution of each structure is assumed to be Gaussian. We run the two algorithms, \texttt{Mclust} and {\sc{LocKE}}, on the whole simulated sample.

\begin{figure*}
\centering
\subfloat[]{\label{locsubs}\includegraphics[width=\columnwidth]{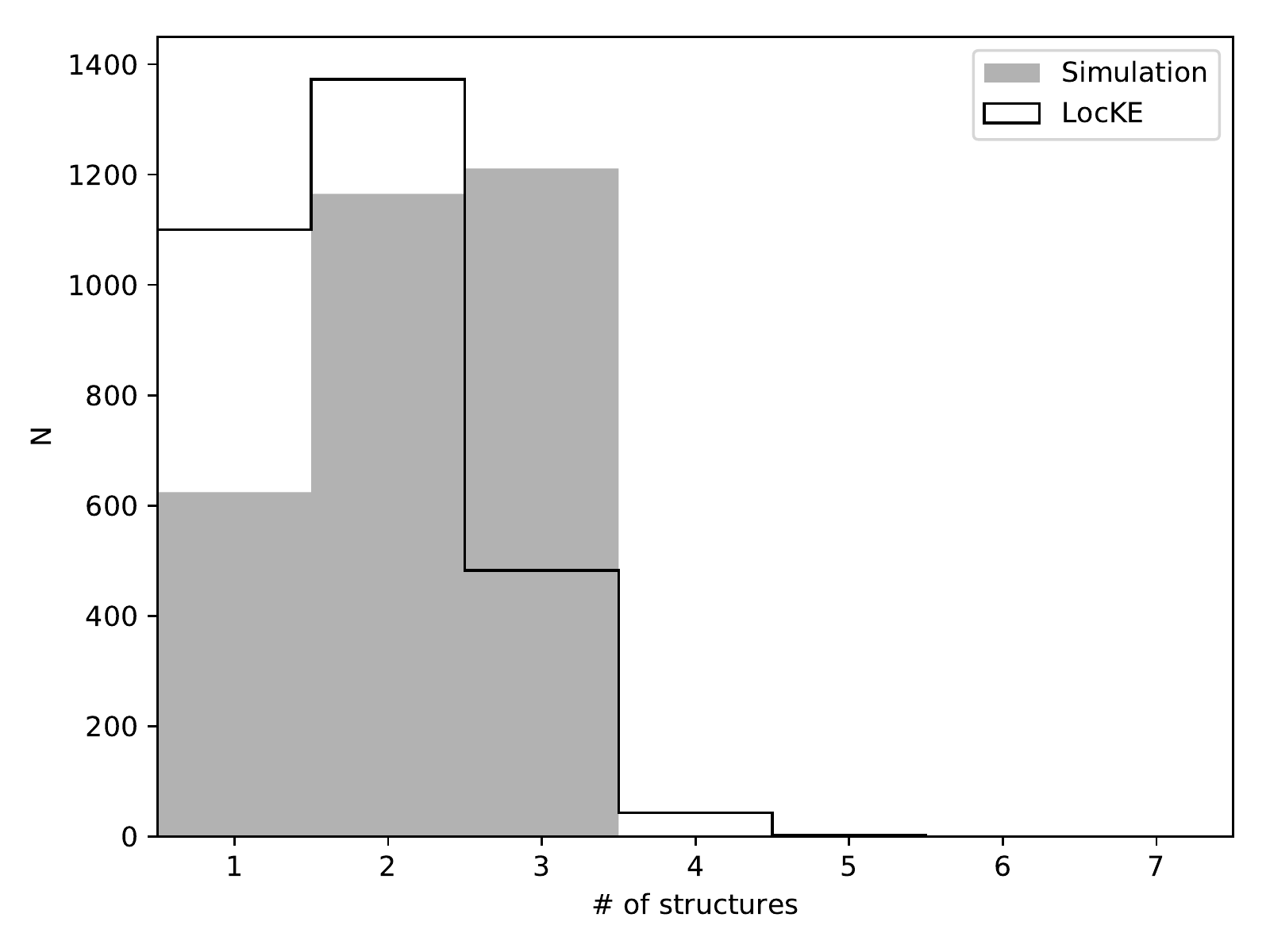}}
\subfloat[]{\label{mclusubs}\includegraphics[width=\columnwidth]{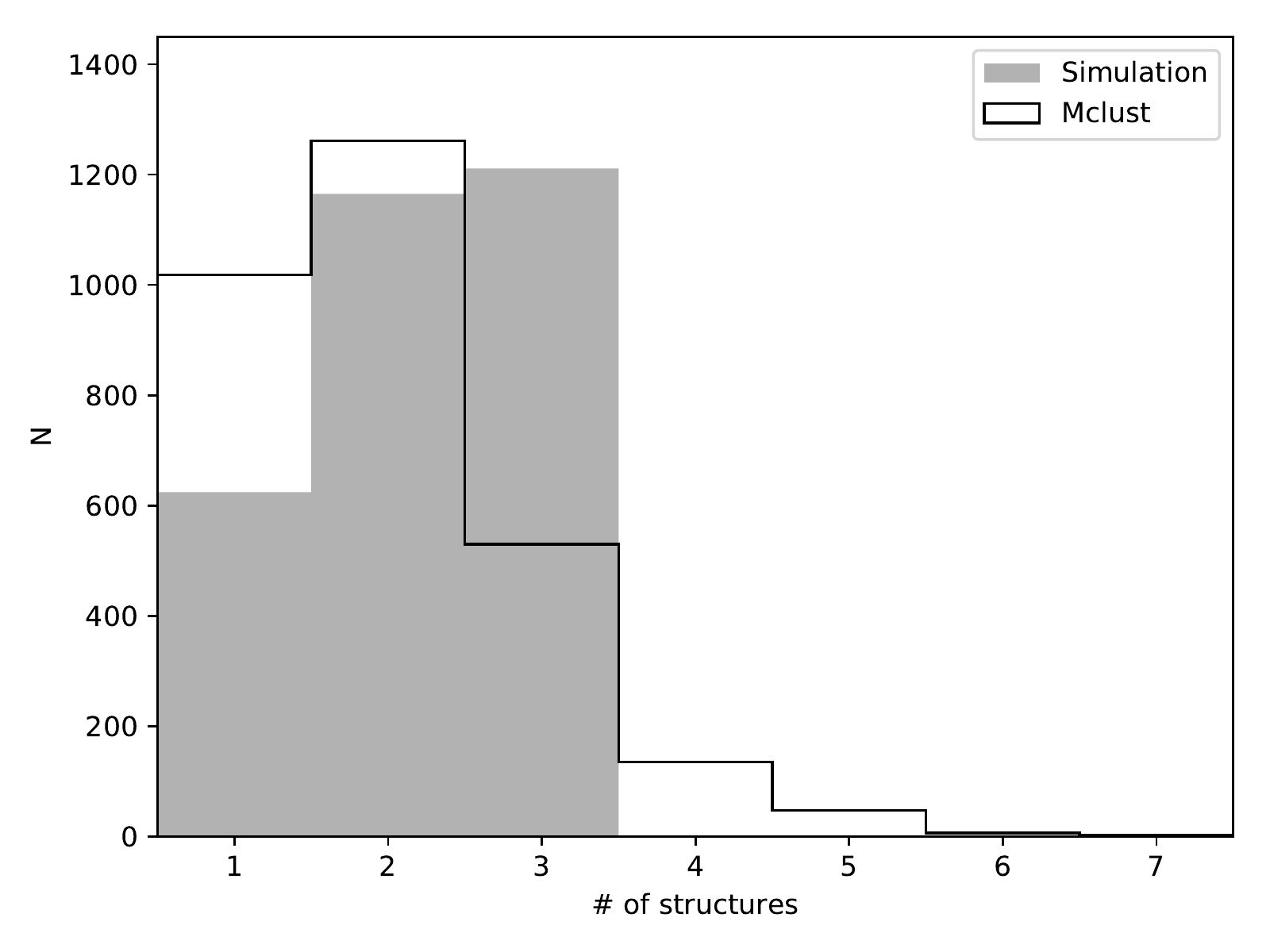}}
\caption{Comparison between the distributions of the input number of structures in the simulated clusters and the number of structures detected by {\sc{LocKE}} and \texttt{Mclust}. The vertical axes display the absolute number ($N$) of simulations with one, two or three substructures.}
\label{num-subs-locke-mcluster}
\end{figure*}

The frequency of substructures found by \texttt{Mclust} (66\%) was similar to that found by {\sc{LocKE}} (63\%); however \texttt{Mclust} found more clusters with 4 or more levels of structures than {\sc{LocKE}}, as seen in the histograms in Figure \ref{num-subs-locke-mcluster}. Both methods fail to completely identify the simulated substructures. It was expected that the detection efficiency would not be 100\% for both codes because, for any algorithm, it is not an easy task to find a substructure when its kinematic parameters are comparable to those of the primary component, and this difficulty is still more pronounced when the structures are much poorer than the primary. This difficulty in reproducing the simulated data can be seen in the histograms of Figure~\ref{num-subs-locke-mcluster}. Around 50\% of all simulated clusters made up of three kinematical structures are not detected as such -- they are mainly classified as a single or a binary cluster but, especially by \texttt{Mclust}, can also be attributed more than the original three structures. We have therefore a combined effect of under-detection of smaller structures and false detection of structures, which actually produce a partial randomization of the detected number of structures with both codes. Regarding the false detection of substructures, however, {\sc{LocKE}} does a much better job: among the 624 simulated clusters without substructures, {\sc{LocKE}} correctly identified 560 ($\sim$90\%) as such, while \texttt{Mclust} identified as such only 319 ($\sim$51\%). We can not define formal completeness limits at this stage, because a successful detection of a substructure involves also the correct identification of their spatial location and kinematics.

A comparison between the simulated and measured properties of each structure is non-trivial. Any detected structure ideally corresponds to a specific simulated structure, but the correspondence between them has to be assessed using some kind of metric. One could use the number of members in each structure, or the kinematic parameters, or a combination thereof. For simplicity, we have chosen an euclidean metric in the mean radial velocity -- velocity dispersion space. Detected structures have been assigned to the simulated ones by minimizing this metric using for the mean radial velocity of the detected structure twice the weight of its velocity dispersion. This is a convenient metric, for it depends only in parameters which are part of the standard {\sc{LocKE}} output; we have also tested other, more complex metrics including the number of galaxies per substructure and the substructure centroids, and the results have not changed significantly, so that the discussion below is robust.

\begin{figure*}
\centering
\includegraphics[width=\textwidth]{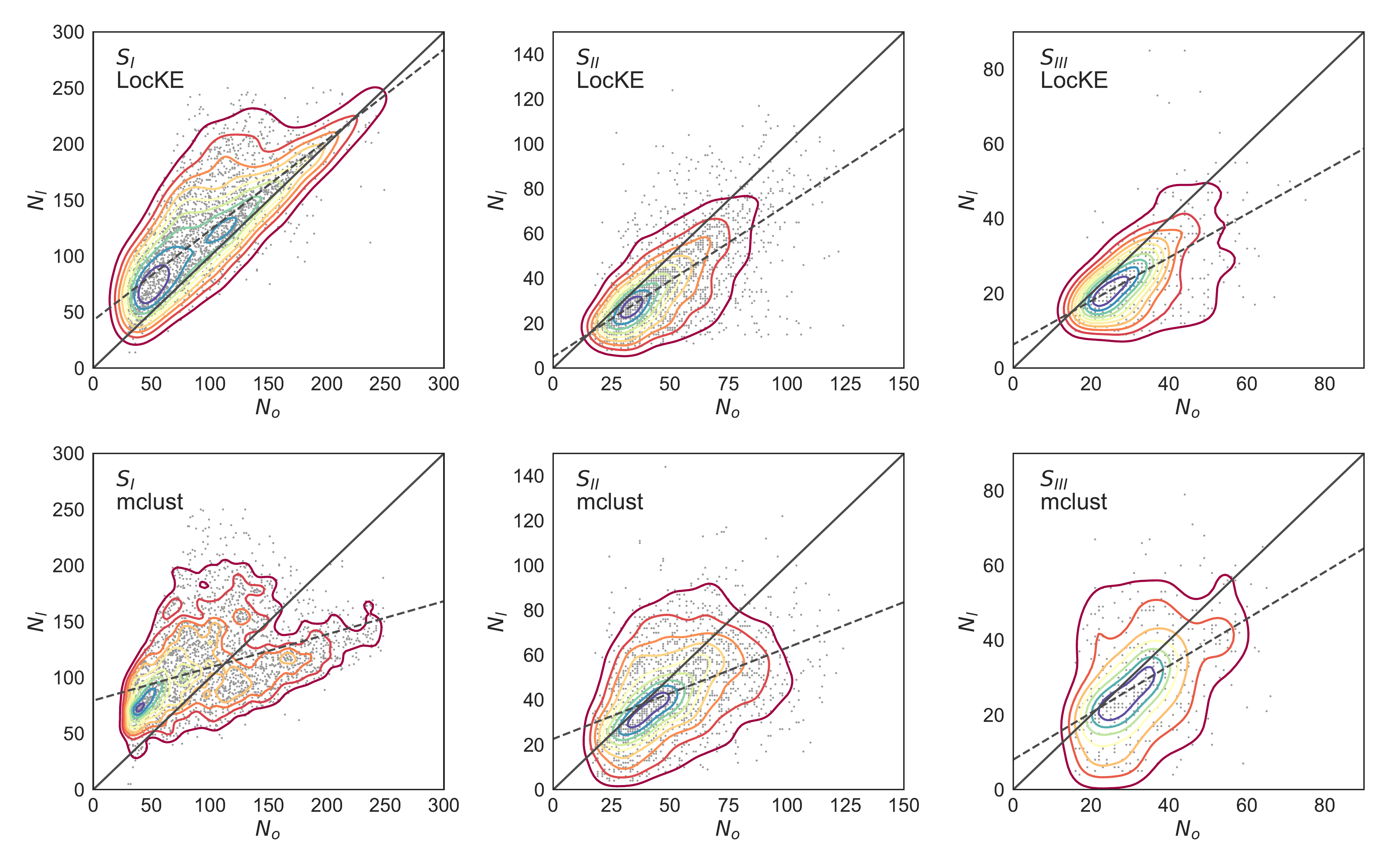}
\caption{Comparison between the simulated number of members in each structure and those derived by {\sc{LocKE}} and \texttt{Mclust}, for primaries ($S_I$), secondaries ($S_{II}$) and tertiaries ($S_{III}$). The vertical axes display the input number of members in each structure, $N_{I}$, and the horizontal axes, the detected number of members $N_{o}$ with each algorithm. Individual structures are represented by gray dots, and the local density of points is represented by contours. The black dashed lines represent the best-fit robust linear regression (see text for details).}
\label{numgals}
\end{figure*}

The comparison between the simulated and the measured number of members attributed to each structure by the two methods can be seen in Figure~\ref{numgals}. For primaries, {\sc{LocKE}} presents a tendency for overestimating the number of neighbours. This is mainly an effect of incompleteness when detecting all substructures present. 
For secondary and tertiary structures, the bias in the number of members is much lower, but the dispersion around the correlation is larger. The Spearman correlation coefficient for primary, secondary and tertiary structures are $0.63$, $0.57$ and $0.47$ respectively. Regarding the slope of these correlations, a Huber robust linear regression (the black dashed lines in Figure~\ref{numgals}) performed with the {\sc{sklearn.linear\_model}} package results in slopes of $0.81$, $0.68$ and $0.58$ for primaries, secondaries and tertiaries respectively, confirming that the correlation is close to the 1:1 relation for primaries, but less so to smaller structures. For \texttt{Mclust}, primaries display a bimodal distribution which is not seen in the {\sc{LocKE}} results. A ``wing'' of fewer detected members can be seen, suggesting that a significant number of galaxies in the primaries are lost to smaller structures. Also, the dispersion in the distributions are consistently higher. The Spearman correlation coefficients for primaries, secondaries and tertiaries, with \texttt{Mclust}, are $0.47$, $0.44$ and $0.44$, and the slopes of these distributions
are $0.30$, $0.41$ and $0.63$.
Notice that, in Figure~\ref{numgals}, we display only the results for simulated clusters with more than one structure, because the number of galaxies detected will be, by construction, the same as the input number for this kind of simulated cluster. For simulated clusters without substructures
and for which {\sc{LocKE}} or \texttt{Mclust} detected at least one substructure, we have calculated the fraction of galaxies which are correctly assigned to the main (richest) structure in the cluster. In this situation, \texttt{Mclust} recovers 63\% of the number of galaxies in the main component, while {\sc{LocKE}} presents
a much higher rate of 83\%.

\begin{figure*}
\centering
\includegraphics[width=\textwidth]{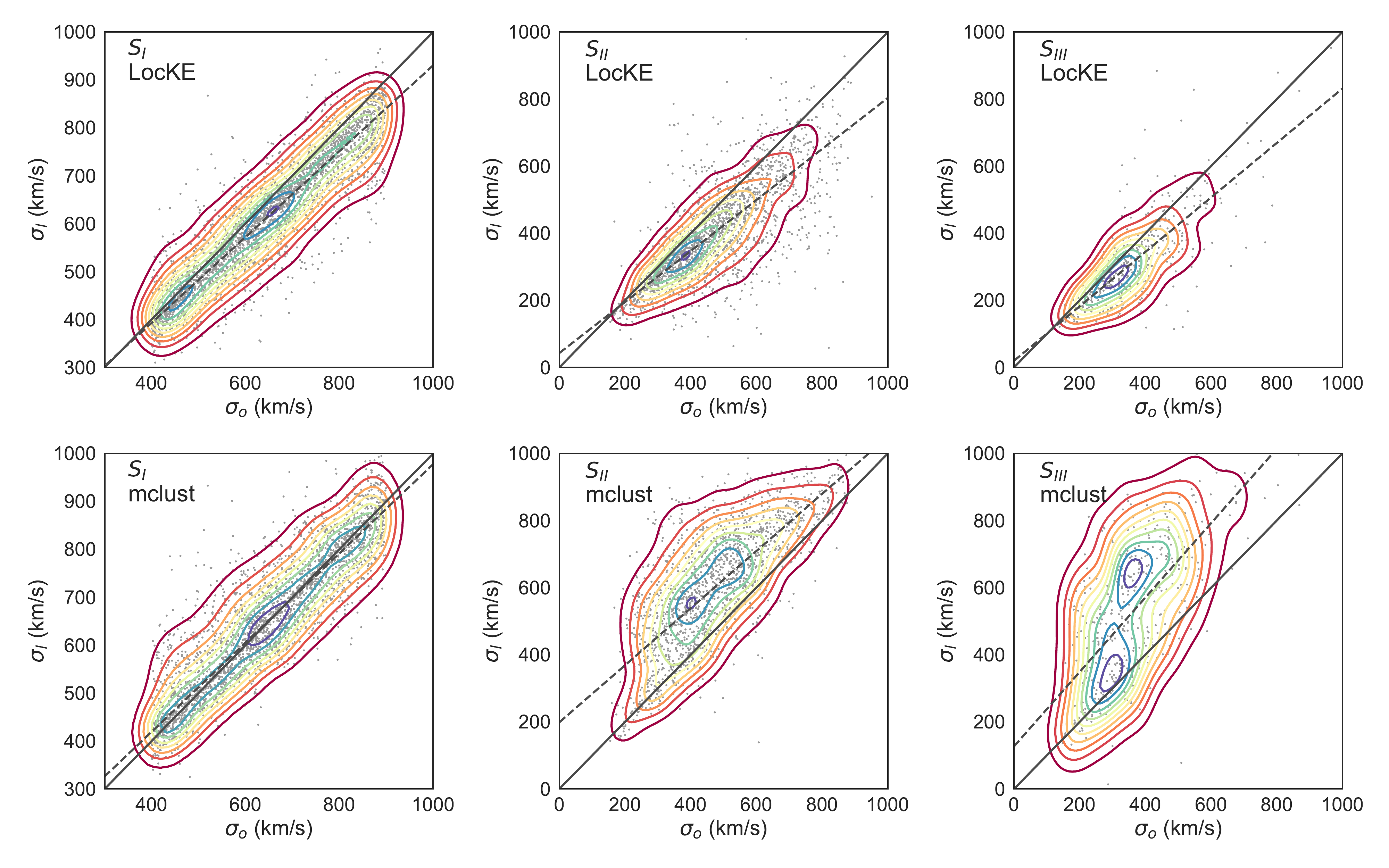}
\caption{Comparison between the simulated velocity dispersion and those measured with {\sc{LocKE}} and \texttt{Mclust}, for primaries ($S_I$), secondaries ($S_{II}$) and tertiaries ($S_{III}$). Symbols are the same as in Figure~\ref{numgals}.}
\label{dispvel}
\end{figure*}

In Figure~\ref{dispvel} we present the comparison between the simulated and measured velocity dispersion using both algorithms. For primaries, both codes perform comparably well; the Spearman correlation coefficient are $0.92$ and $0.90$ for {\sc{LocKE}} and \texttt{Mclust} respectively, and the respective slopes are $0.90$ and $0.93$. However, for secondaries and tertiaries, \texttt{Mclust} introduces a very
strong bias in the sense of overestimating the simulated velocity dispersion of the substructures. This is a result from the same contamination by galaxies in the primary
component which produces the upper ``wing'' in Figure~\ref{numgals}. {\sc{LocKE}}, on the
other hand, does not bias the velocity dispersion estimates for secondaries and tertiaries.
The Spearman correlation coefficients for secondaries and tertiaries are $0.75$ and $0.77$ for {\sc{LocKE}} and $0.71$ and $0.65$ for \texttt{Mclust}.
The slopes of the correlation for secondaries with {\sc{LocKE}} and \texttt{Mclust}
are 0.76 and 0.85, but the intercept is much larger for \texttt{Mclust} ($198.5$\,km/s) than for {\sc{LocKE}} ($42.9$\,km/s). A similar
behavior is observed for tertiaries -- slope $0.81$ ($1.11$) and intercept $19.9$\,km/s ($126.8$\,km/s) for {\sc{LocKE}}
(\texttt{Mclust}).

\begin{figure*}
\centering
\includegraphics[width=\textwidth]{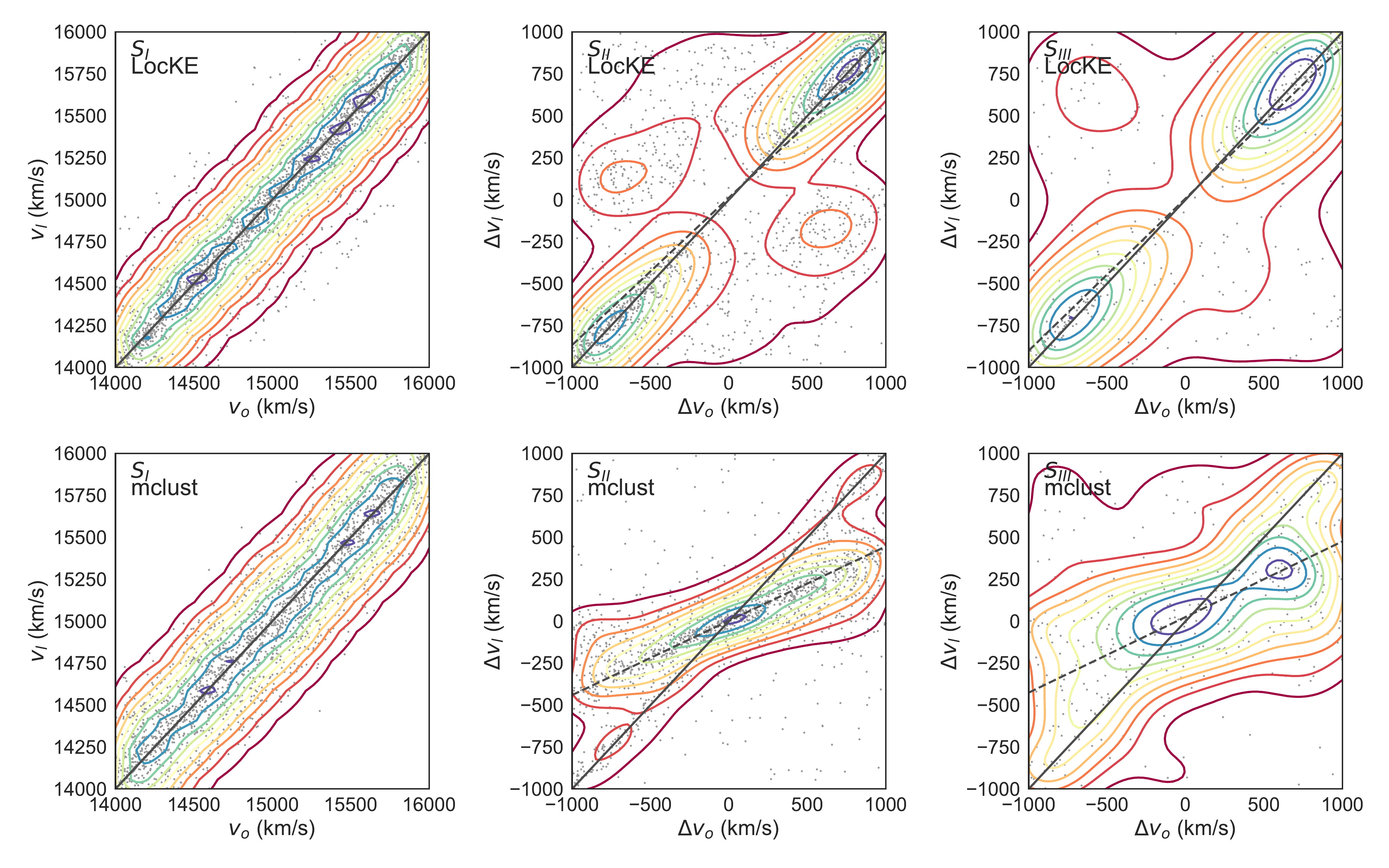}
\caption{Comparison between the simulated mean line of sight velocity and those measured with {\sc{LocKE}} and \texttt{Mclust}, for primaries ($S_I$), secondaries ($S_{II}$) and tertiaries ($S_{III}$).  Symbols are the same as in Figure~\ref{numgals}.}
\label{vel}
\end{figure*}

Finally, we show in Figure~\ref{vel} the mean line-of-sight velocities of the simulated
and detected structures. For primaries, the correlation is excellent for both methods (Spearman correlation coefficients of $0.95$ and $0.93$ for {\sc{LocKE}} and \texttt{Mclust} respectively) and no significant bias is observed. The performance of \texttt{Mclust} for secondaries and tertiaries, however, shows a pronounced
bimodality: a family of solutions defining the bi-sector of the
correlation plot, plus a less steep, nearly-horizontal distribution. This behaviour
is once again due to contamination from galaxies in the primary structure, that tend to
reduce the relative line-of-sight velocities (defined as zero in the reference
of the primary). {\sc{LocKE}} on the other hand presents a better behaved relation,
with some scattering outside the bi-sector and mostly due to a wrong assignment of the 
substructure level (primary/secondary). The correlation coefficients for secondaries and tertiaries are $0.66$ and $0.60$ for {\sc{LocKE}} and $0.74$ and $0.52$ for \texttt{Mclust}. (Notice that, even though the correlation coefficient for secondaries
is larger for \texttt{Mclust} than for {\sc{LocKE}}, the slope of the correlation
is strongly biased.)

We therefore conclude that {\sc{LocKE}}, even though presenting the intrinsic weaknesses
of the $\delta$ and $\kappa$ tests (the inability of identifying substructures aligned in the line
of sight), it does a good job at detecting and estimating the kinematic properties of substructures,
and also in isolating their members. Though introducing some level of bias in these parameters, 
particularly due to undetected structures, they are much better behaved than multidimensional
mixture as modelled by \texttt{Mclust}, while maintaining its main strength of being fully automated.
We have used {\sc{LocKE}} to investigate the presence of substructures in our sample 412 selected clusters of galaxies and present the results in Sect.~\ref{sect:freqs}.

\subsection{Stellar population synthesis}
\label{sect:sps}

To investigate the influence of the dynamic stage of the cluster in the stellar populations of galaxies, we performed stellar population synthesis and derived the mean stellar age of each galaxy in the sample. For this, we made use of the {\sc{starlight}} code \citep[]{cid2005,mateus2006}. As templates
we have used single stellar population models (SSPs) from the Medium resolution Isaac Newton telescope Library of Empirical Spectra (MILES) \citep[]{sanchez2006}, which match roughly the SDSS
spectral resolution ($\sim$2.3\AA) and cover the optical range from 3525 to 7500 {\AA}. We have
selected a set of 156 SSPs with 26 different ages ranging from 70.8\,Myr up to 14.12\,Gyr, and 6 different metallicities ranging from $Z=0.0004$ up to $Z=0.0315$ (i.e.
from $Z=0.02Z_\odot$ to $Z=1.575Z_\odot$). Prior to the synthesis, the
observed SDSS spectra have been converted to linear scale, resampled to steps of 1\AA{}, corrected for Galactic extinction and corrected to restframe with the Python code {\sc{pystarlight}} \footnote{Available to download at \emph{http://www.starlight.ufsc.br/node/3}}.

The {\sc{Starlight}} output includes the mass and light contribution of each SSP to the observed
spectrum. The light-weighted mean stellar age $\bar{t}$ of a galaxy is defined by

\begin{equation}
\log \bar{t} = \displaystyle \sum_{j=1}^{156} \log t_{j} L_{j} / \displaystyle \sum_{j=1}^{156}L_{j},
\label{agecalc}
\end{equation}

\noindent where $t_j$ is the age of template $j$ and $L_j$ is its contribution to the observed spectrum. We have also derived the stellar masses, in units of
the solar mass, using

\begin{equation}
M_{\star}=\frac{M_{cor\_tot} \times 10^{-17} \times 4\pi dl^{2}}{L_{\odot}},
\label{mstar}
\end{equation}

\noindent where $M_{cor\_tot}$ is a {\sc{Starlight}} output, $dl$ is the luminosity distance, and $L_{\odot}$ is the solar luminosity. We then correct the fiber stellar mass obtained above to total stellar mass by scaling the total and fiber $r$-band magnitudes, resulting in an increase of 4\% on average relative to the fiber stellar masses.

\section{Results}

\subsection{Frequency of substructures}
\label{sect:freqs}

For the full sample of 408 clusters presenting more than 30 members, {\sc{LocKE}} detected substructures in 180, or $\approx$ 45\% of the sample. This value is in agreement with typical frequencies using a single method for substructure detection. \citet[]{Aguerri2010}, for example, 
using the $\delta$ test, have found substructures in 34\% of their sample of 88 rich galaxy clusters.
\citet[]{Solanes1999} studied a sample of 67 rich clusters extracted from ENACS catalog using four different statistical tests. A normality test indicated that 30\% of the sample are unrelaxed, and with the $\delta$-test they found a frequency of substructures of 31\%. \citet[]{Einasto2012},
on the other hand, adopted multiple 1-D (Shapiro-Wilk and Anderson-Darling), 2-D ($\beta$-test) and
3-D ($\alpha$-test, $\delta$-test and multidimensional mixture modelling with \texttt{Mclust}) to investigate 109 clusters of galaxies, and found a large frequency of 70\% (80\%) of clusters
with substructures as indicated by the $\delta$ test (\texttt{Mclust}).

For further analysis, we will split the sample clusters in three broad categories. Clusters
with no substructures detected will be referred to as ``no substructures'' and represented by NS.
For clusters presenting substructures, we will refer separately to the main structure, or main halo, defined as the most massive structure in the cluster (represented by PC, or ``Primary Component'') and all lower mass structures as ``secondaries'' no matter the number of structures detected (hereafter, SC or ``Secondary Component''). We will also refer to the full population of galaxies in clusters presenting substructures, before separation in individual units by {\sc{LocKE}}, as ``unrelaxed'', and represent them by the symbol U, when convenient.

\begin{figure}
 \includegraphics[width=\columnwidth]{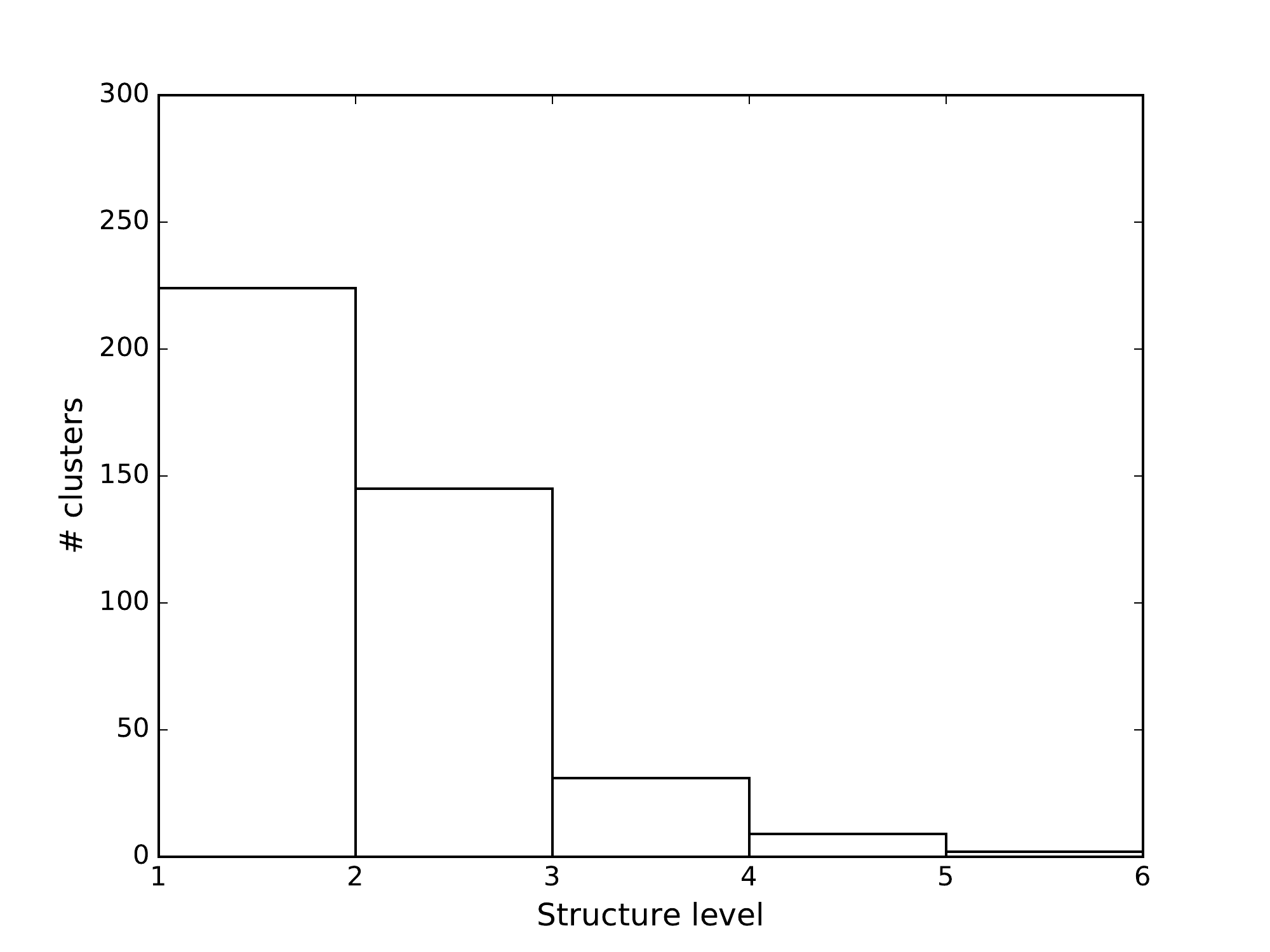}
 \caption{Number of clusters of galaxies in our sample as a function of the number of kinematic structures identified by {\sc{LocKE}}.}
 \label{stlevel}
\end{figure}

\subsection{Properties of the individual kinematic components}

\begin{figure}
 \includegraphics[width=\columnwidth]{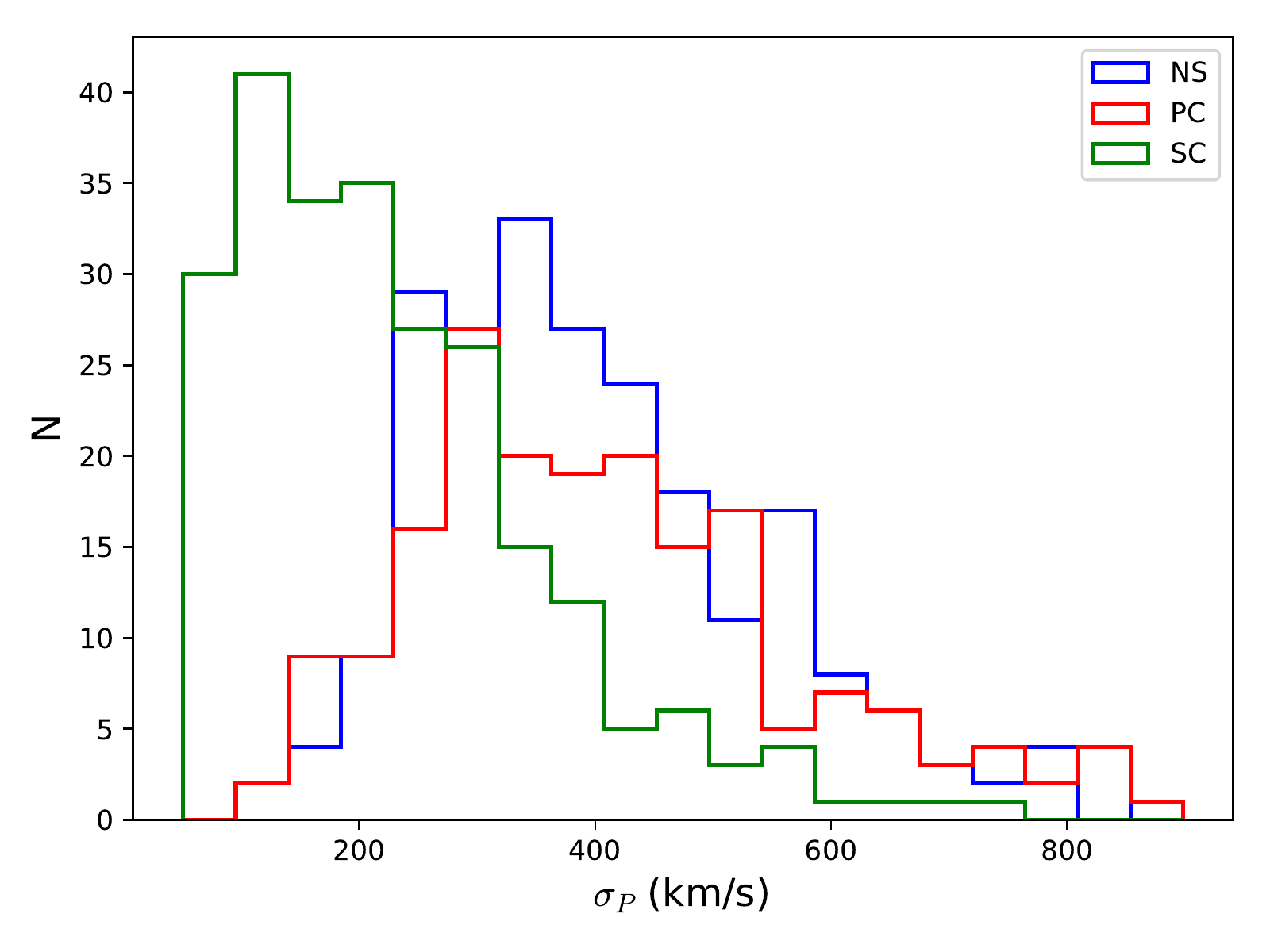}
 \caption{Velocity dispersion distribution for clusters without substructure (NS), primary components (PC) and secondary component (SC).}
 \label{disp1}
\end{figure}

After the identification of kinematic substructures, {\sc{LocKE}} provides kinematic parameters for each structure. Of particular importance is the velocity dispersion, that is used to calculate the virial mass. In Figure~\ref{disp1} we show the velocity dispersion distribution of all structures
found in our sample. The blue, red and green curves represent NS, PC and SC clusters; the mean velocity dispersion for each class is $400\pm 76$\,km/s, $417\pm 96$\,km/s and $231\pm 81$\,km/s. These values are typical of medium-sized clusters and groups \citep[e.g.]{Colles1996,Aguerri2010,hou2012}.

\begin{figure}
 \includegraphics[width=\columnwidth]{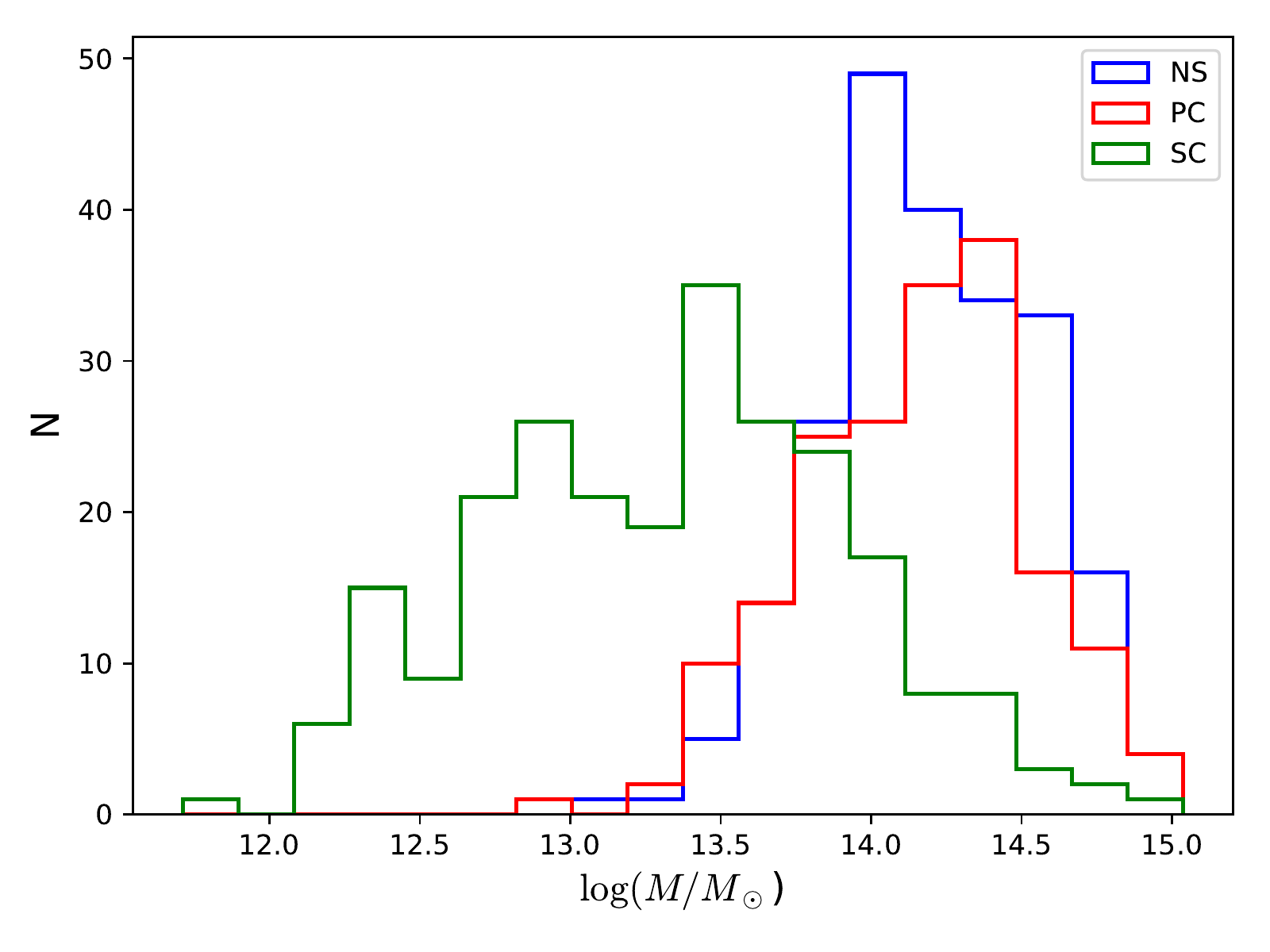}
 \caption{Mass distributions for NS, PC and SC structures of the clusters in our sample. Masses of PC and NS structures are similar, and larger than the typical masses of the secondary components.}
 \label{mass1}
\end{figure}

\begin{figure}
 \includegraphics[width=\columnwidth]{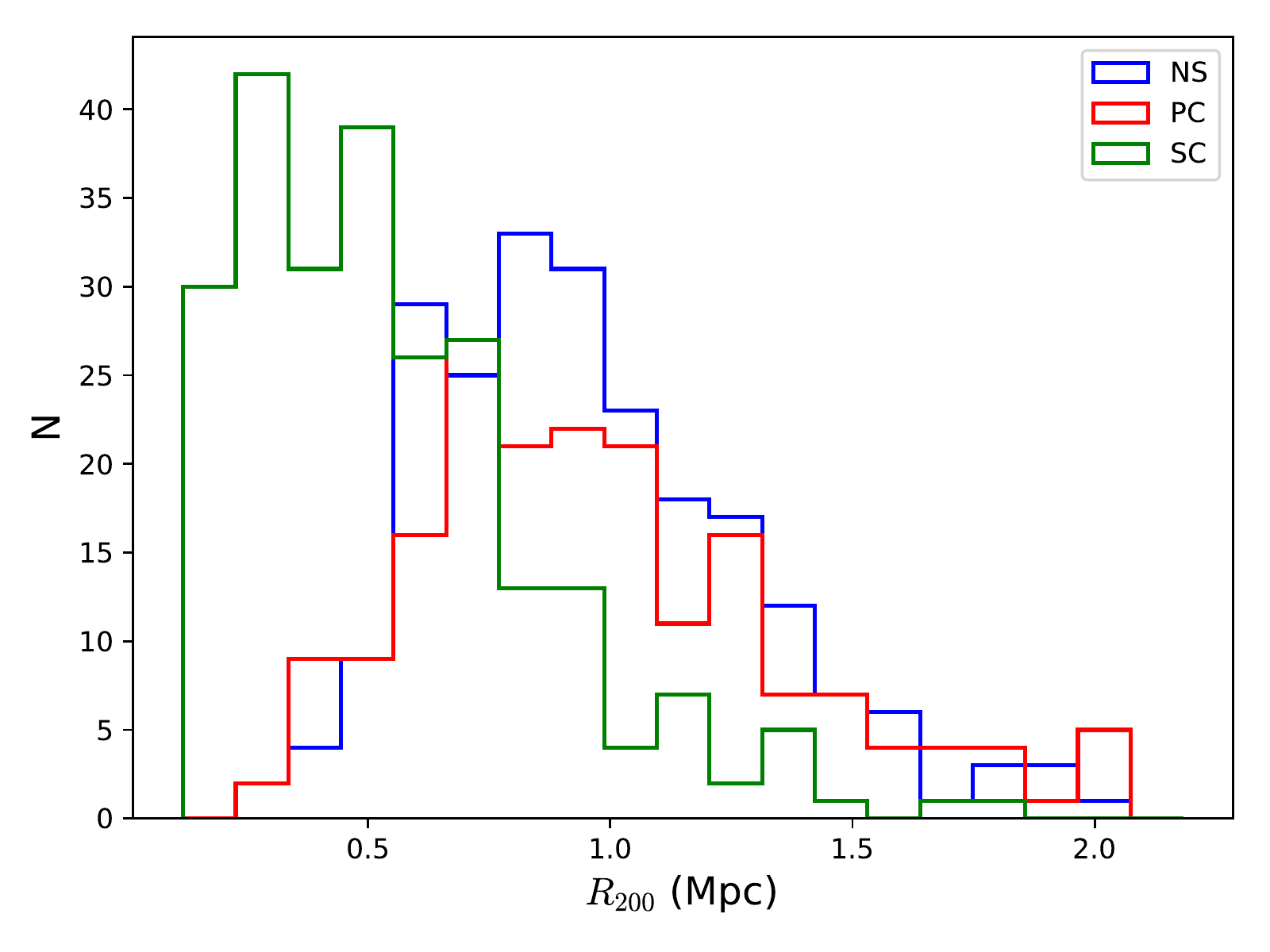}
 \caption{Distribution of $R_{200}$ values for all structures in our sample, showing similar sizes for NS and PC structures; SC structures tend to be much smaller.}
 \label{r200his}
\end{figure}

Figure~\ref{mass1} presents the mass distribution of all structures of the sample as derived by 
{\sc{LocKE}}. The mean values for the logarithmic virial mass are $14.19$, $14.17$ $13.34$ dex for NS, PC and SC, respectively, which are also in agreement with typical values for
this class of object \citep[e.g.]{carlberg1996,Lopes2009}. Regarding the physical size of each structure, we have found $R_{200}$ values of 0.96\,Mpc, 1.00\,Mpc and 0.55\,Mpc for NS, PC and SC, respectively. The distribution of this parameter is shown in Figure~\ref{r200his}, also in agreement with other authors \citep[e.g.][]{carlberg1997,Lopes2009,yan2015}.

The above distributions show that primary structures (PC) are comparable in mass and radius to
the NS class. These two classes are therefore physically similar, with the exception that
the former are associated to other nearby structures. Secondaries, on the other
hand, are consistently less massive, and smaller in size than PC and NS.

\subsection{Stellar populations in individual structures}

\begin{figure*}
 \includegraphics[width=\columnwidth]{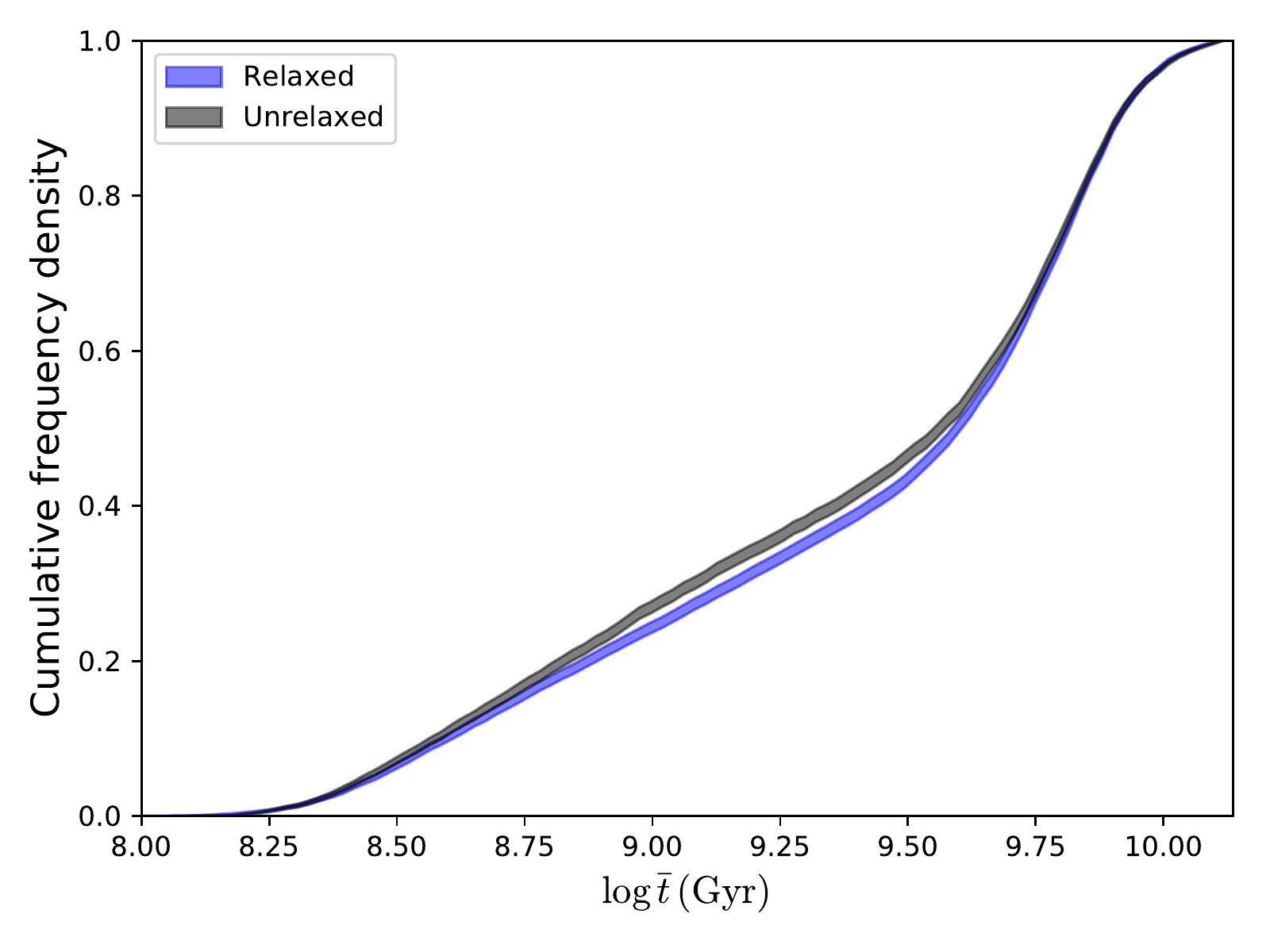}
 \includegraphics[width=\columnwidth]{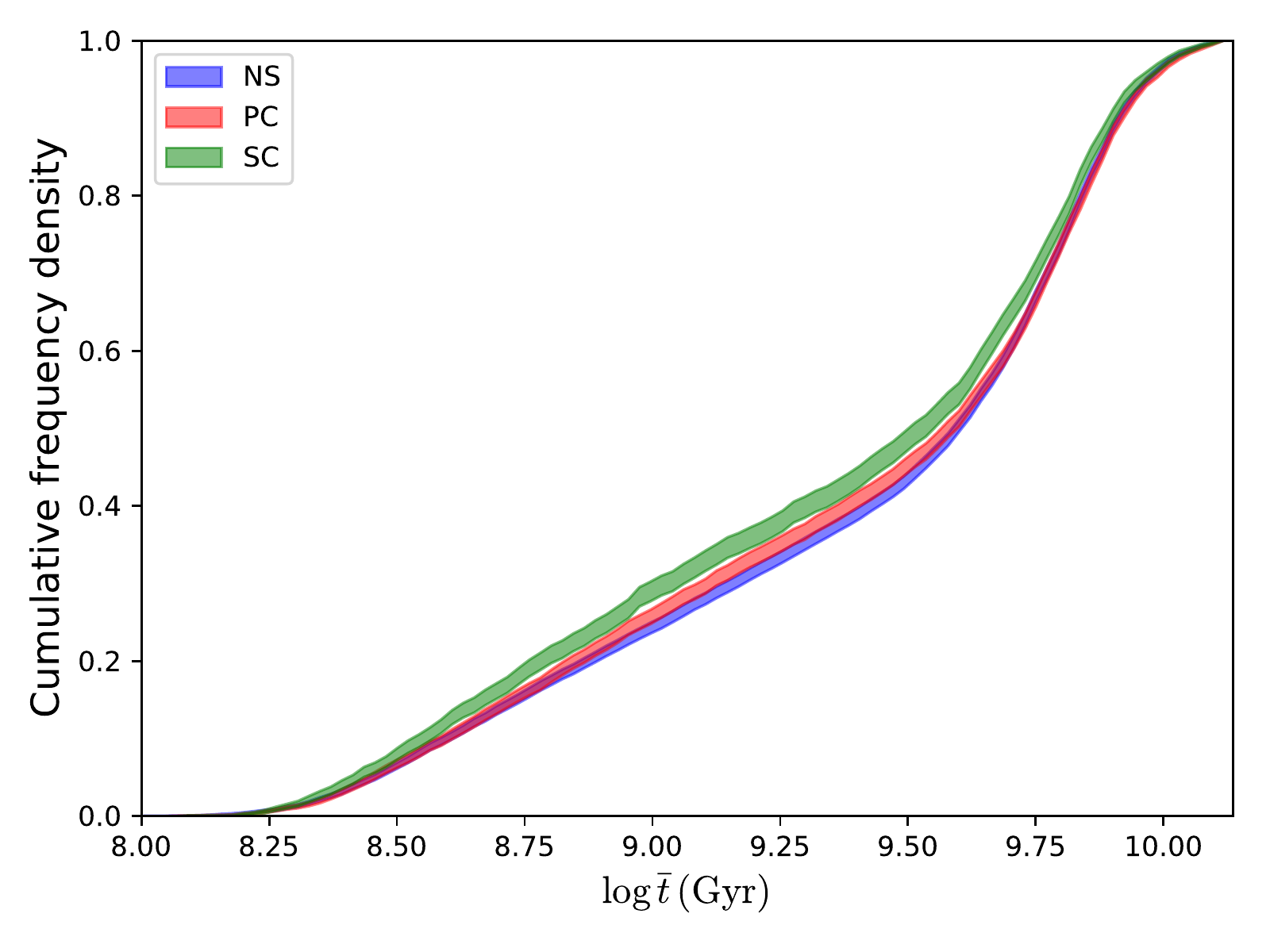}
 \caption{Cumulative frequency of the logarithmic stellar age distribution for galaxies in each structure class defined in this work. In the left panel, relaxed clusters are represented by the blue shade, and unrelaxed (U) clusters are shown in black. In the right panel, the blue, red and green curves represent NS, PC and SC structures, respectively.}
 \label{agehis}
\end{figure*}

In this section we investigate how the typical mean stellar ages of galaxies
in our sample respond to the class of structure they populate. Figure~\ref{agehis} presents the cumulative distributions of the light-weighted logarithmic mean stellar age of galaxies in each structure class. The thickness of the curves indicate the poisson errors. We have chosen to present the cumulative distributions instead of a discrete histogram in order to make the differences more clear. The age distribution of relaxed (NS) versus unrelaxed (U) clusters are distinct, with unrelaxed clusters (in black) presenting a larger contribution of galaxies with younger stellar populations. When dividing U clusters into primaries and secondaries, the behavior of the mean stellar age is quite similar among the three classes, with some subtle differences. Galaxies located in the secondaries (SC; green curve) tend to present younger stellar populations. For galaxies in the primaries (PC; red curve) ages tend to be slightly lower than for NS (blue curve) clusters, but the difference is very small. This means that, when excluding the lower mass substructures from unrelaxed clusters, the resulting age distribution becomes much closer to the relaxed cluster sample.

\begin{figure}
\includegraphics[width=\columnwidth]{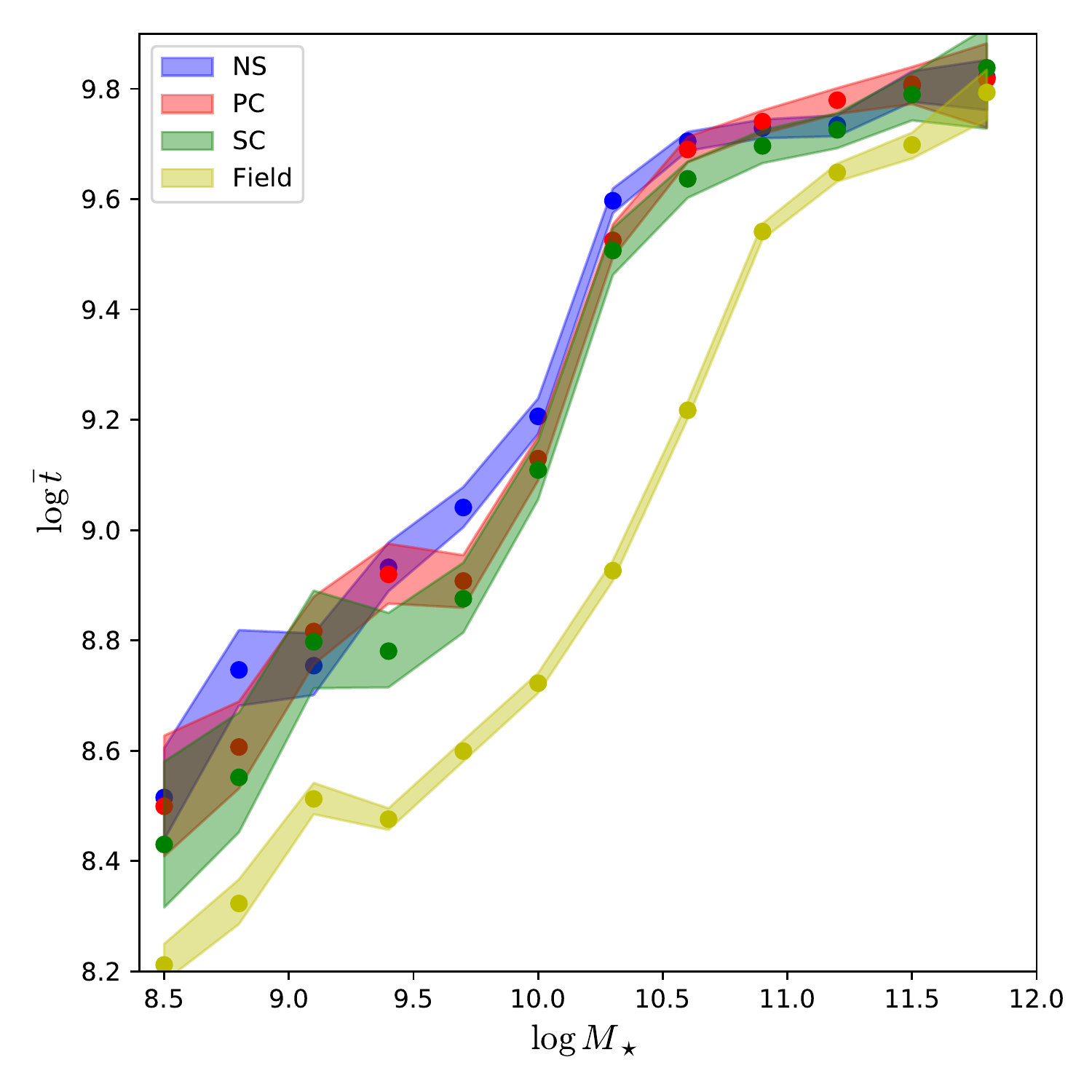}
\caption{Mean stellar age as a function of stellar mass for galaxies in each of the three structure classes defined in this work and for a sample of field galaxies. The blue, red and green curves represent NS, PC and SC structures, respectively, and yellow curve represent the field sample.}
\label{ss-cp-cs}
\end{figure}

Trying to clarify these differences in mean stellar age, we now investigate how these depend on galaxy stellar mass. In Figure~\ref{ss-cp-cs} we present the mean stellar ages for our three classes of structures as a function of their stellar mass. The error bars have been obtained via bootstrap resampling with 10,000 realizations. The most evident feature of this figure is that galaxies with higher stellar mass are older for all classes, including field galaxies (yellow curve). This reflects the well-established phenomenon of \emph{downsizing}, in which the star formation history is skewed towards lower redshifts for less massive galaxies \citep[]{neistein2006,fontanot2009}. We also can see that galaxies in secondary structures (SC) seem to present lower mean ages relative to primary components (PC) and clusters with no substructure (NS). These differences are very small -- corresponding to less than 1\,Gyr in mean stellar age -- and stronger at masses lower than $10^{10.8}\odot$. At the lowest mass bins ($\lesssim 10^9\odot$), the difference is less clear due to the large statistical noise. Galaxies in PC also tend to present younger stellar populations relative to NS clusters, but the difference is lower than for SC at all mass bins. Field galaxies are younger, for all mass bins, than galaxies in any cluster structure, and this difference can reach up to $\sim 3$\,Gyr.

\begin{figure}
\centering
\includegraphics[width=\columnwidth]{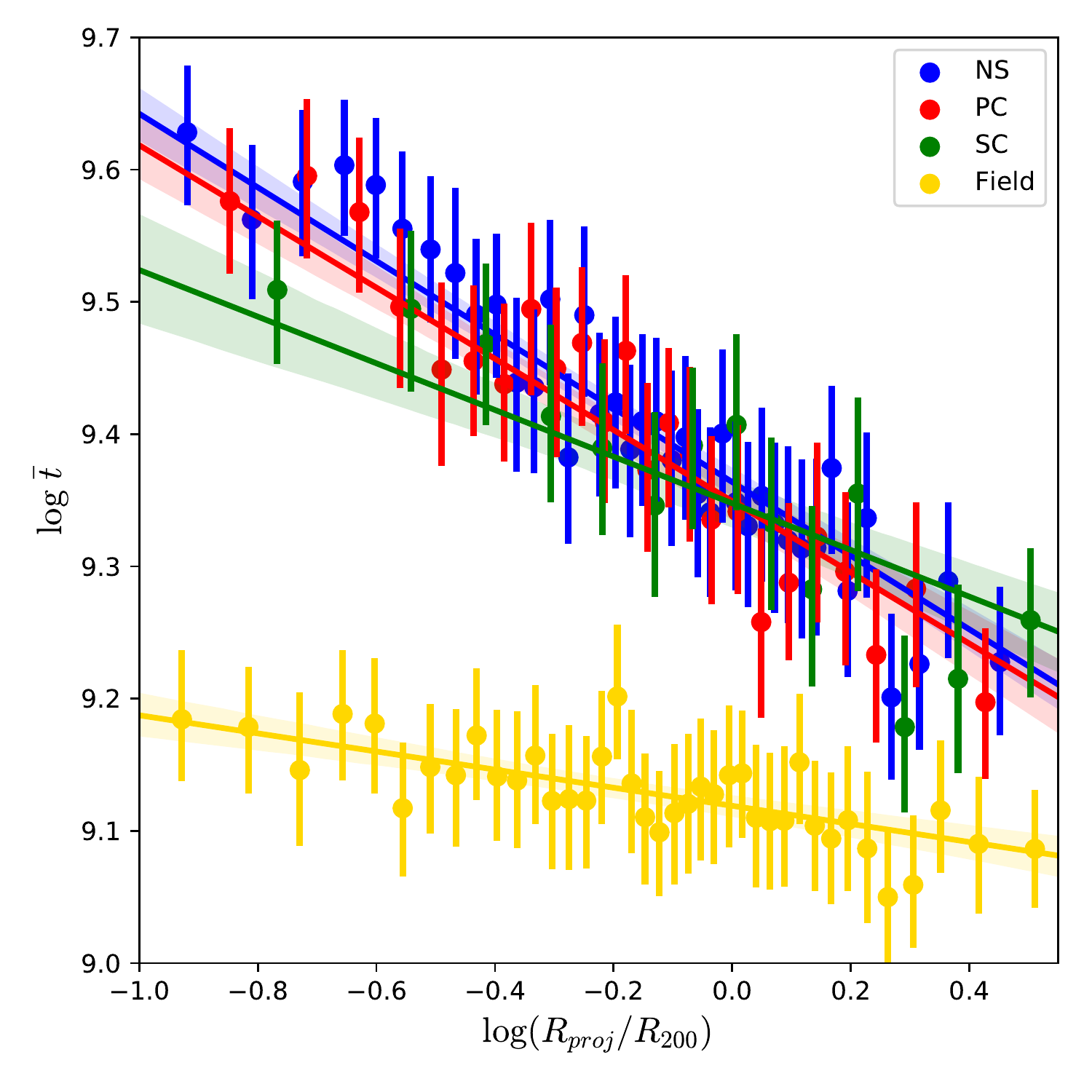}
\caption{Logarithm of mean stellar age as a function of the distance of the galaxies to the center of its parent structure. Each dot represents the average mean stellar age in bins of radial distance. Blue dots represent relaxed (NS) clusters, red (green) dots stand for primaries (secondaries) of unrelaxed clusters, and yellow dots designate field galaxies. The error bars indicate the standard error of the mean.}
\label{clusrad}
\end{figure}

We now investigate how the mean stellar age of galaxies respond to their clustercentric distance. In Figure~\ref{clusrad} we present the mean stellar age as a function of the galaxy clustercentric distance normalized to $R_{200}$, separating the sample according to the three structure classes. We notice an overall tendency of lower mean stellar ages occur at large clustercentric distances for the three classes. This result reflects the well-established relationship between galaxy properties (morphology, age, SF, etc.) and clustercentric distance \citep[e.g.][]{dressler1980,kauffmann04,blanton05}. However, this radial mean age gradient can be seen for \emph{all magnitude ranges and structure classes}, so that galaxies closer to the center of every structure present older stellar populations than galaxies in their outskirts, no matter the status of the structure. This shows that some kind of pre-processing has already been acting on the galaxy population which is being accreted into primaries. We can see, however, that the slope of the age-radius relation is consistently less steep for secondaries.
For primaries, on the other hand, the age-radius relation seems to be completely in place. Using cosmological N-body simulations, \citet[]{ludlow09}, have shown that, during the process of virialization of a cluster after accreting a low-mass galaxy system, the most important parameter defining the orbits of accreted galaxies is the orbital phase coupling between the orbit of the substructure and the main component of the cluster. The dissolution of a substructure is therefore expected to contribute with a new population of galaxies across the full extend of the primary. This in turn implies that a shift in mean stellar age for recently accreted galaxies is expected to occur before the clusters detected as ``unrelaxed'' attain a new virialization state.

Because mean stellar ages are higher for more massive galaxies, it is important to check whether these radial gradients in mean stellar age are an artifact due to a radial gradient in galaxy stellar mass. We have tested this hypothesis in two complementary ways. Firstly, we substitute the mean stellar age of every cluster member (at any structure level) by the mean stellar age of a field galaxy at the same stellar mass. If the observed radial age gradients are produced by a stellar mass gradient, replacing cluster galaxies by field galaxies of the same stellar mass should reproduce the observed age gradients. The result of this experiment is illustrated by yellow dots in Figure~\ref{clusrad}. Some degree of dependence of the mean stellar age on clustercentric radius is indeed observed for galaxies drawn from the field population. However, the resulting gradient is much less steep (by a factor $\sim 6$) than that observed by cluster galaxies in NS and PC structures, indicating that galaxy stellar masses are not the main driver of the observed radial trends in mean stellar age for these massive structures. For SC structures, on the other hand, the slope of this relation is not much larger than that for field clusters. In order to be able to perform a better comparison between the slopes of this relation for cluster galaxies and that obtained with field galaxies, we have split the field sample into three subsamples -- NSField, PCField and SCField -- drawing galaxies following stellar mass distributions similar to each individual structure. We have then performed linear regressions for the derived mean stellar age - clustercentric radius relations for cluster galaxies (NS, PC and SC) and for their corresponding field galaxies (NSField, PCField and SCField). We have obtained the following slopes: $\alpha_{NS} = -0.28 \pm 0.01$, $\alpha_{PC} = -0.27 \pm 0.02$, $\alpha_{SC} = -0.17 \pm 0.02$ for cluster galaxies, and $\alpha_{NSField} = -0.08 \pm 0.01$, $\alpha_{PCField} = -0.07 \pm 0.02$, $\alpha_{SCField} = -0.03 \pm 0.02$ for field galaxies. The differences between these slopes is $-0.20 \pm 0.02$ for both NS and PC structures, and reduces to $-0.14 \pm 0.03$ for SC structures. Therefore, for all structures, both the stellar age gradients and the ``gap'' between slopes of SC structures and more massive ones are shown to be preserved when we take galaxy stellar masses into account.

\begin{figure}
\centering
\includegraphics[width=\columnwidth]{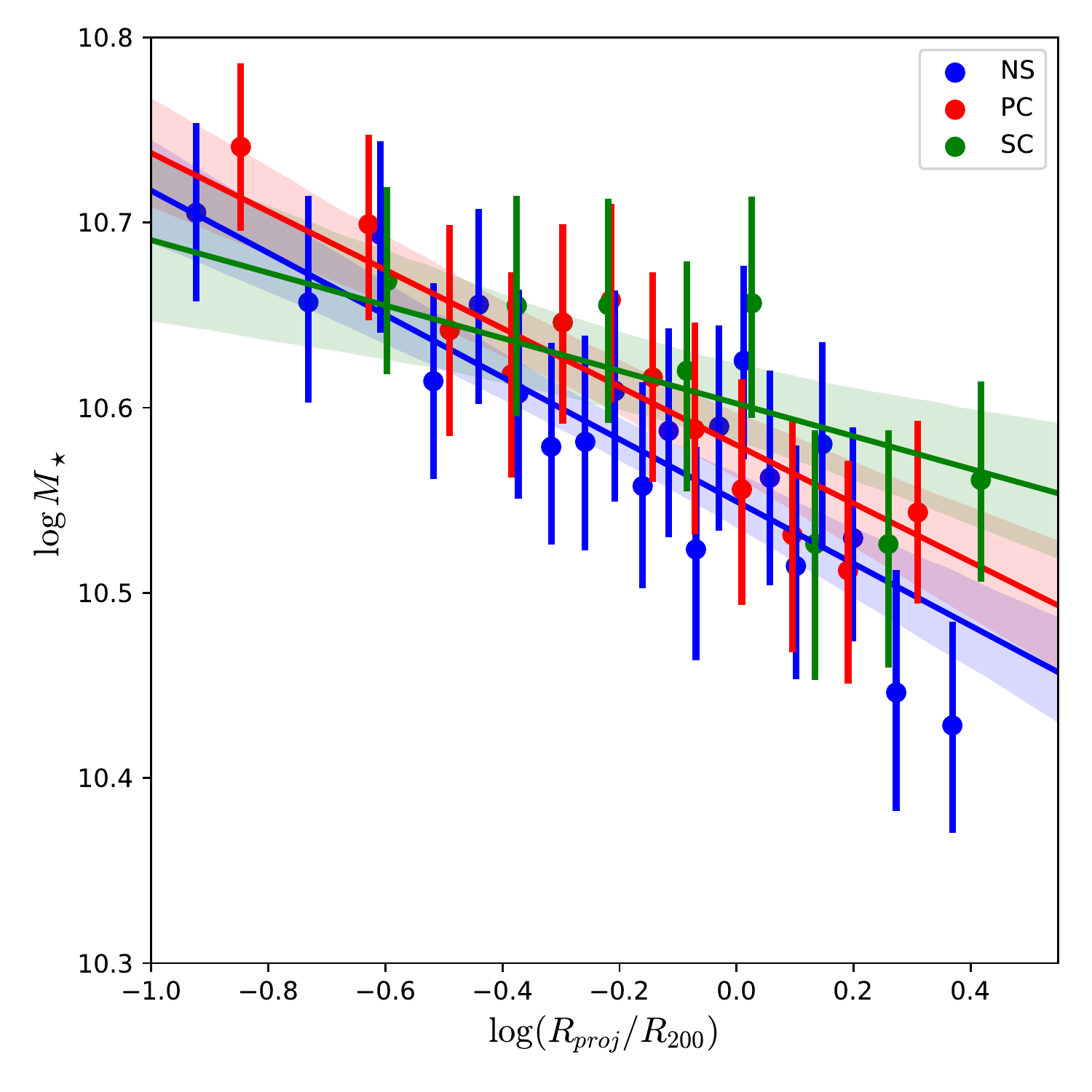}
\caption{Logarithm of stellar mass as a function of the distance of the galaxies to the center of
its parent structure. Each dot represents the average mean stellar age in a bin in radial distance. The dots and colors follow the same configuration as in figure \ref{clusrad}.} 
\label{clusmass}
\end{figure}

The second test we have performed is a direct calculation of the mean stellar mass of cluster galaxies as a function of the radial clustercentric distance. This is presented in Figure~\ref{clusmass}. A radial mass gradient can be detected in our cluster galaxies at all structure classes; in fact, some degree of radial dependence of stellar mass in cluster galaxies has been observed by other authors \citep[e.g.,][]{roberts+15}. However, these gradients are very small -- corresponding to less than 40\% in average stellar mass across the full radial extension of the structure up to $2.5R_{200}$ -- and, more importantly, stellar masses correlate with the clustercentric radius much worse than the mean stellar age. We measured the Spearman correlation coefficient for mean stellar age and clustercentric distance, obtaining $-0.23$ for both NS and PC, and $-0.17$ for SC  structures. The Spearman correlation coefficient between stellar mass and clustercentric distance, in contrast, are $-0.09$, $-0.08$, and $-0.06$ for NS, PC and SC, respectively. It is therefore not possible for these small, disperse clustercentric stellar mass gradients to produce much tighter and steeper mean stellar age gradients.

\begin{figure*}
\centering
\includegraphics[width=\textwidth]{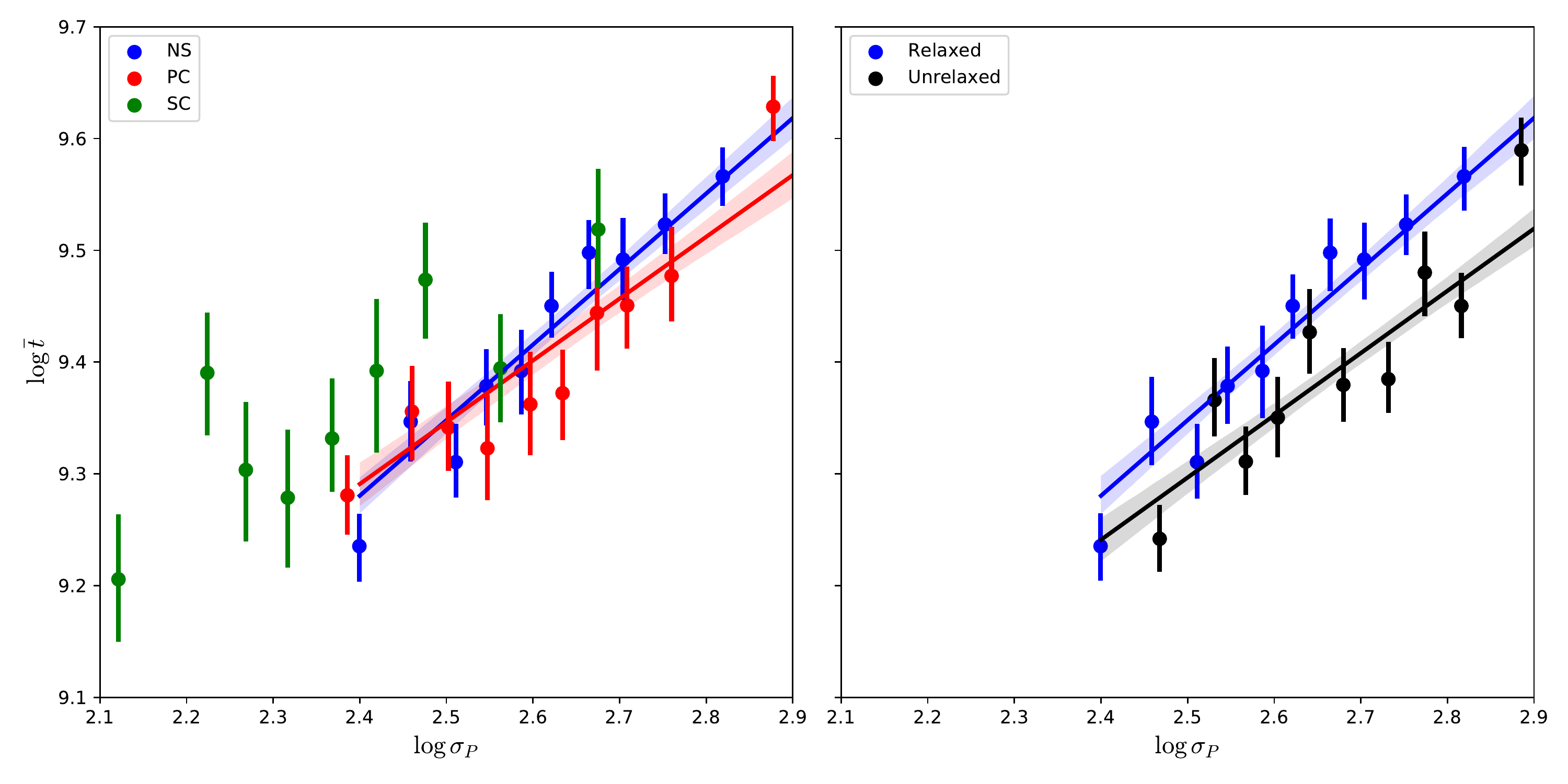} \\
\includegraphics[width=\textwidth]{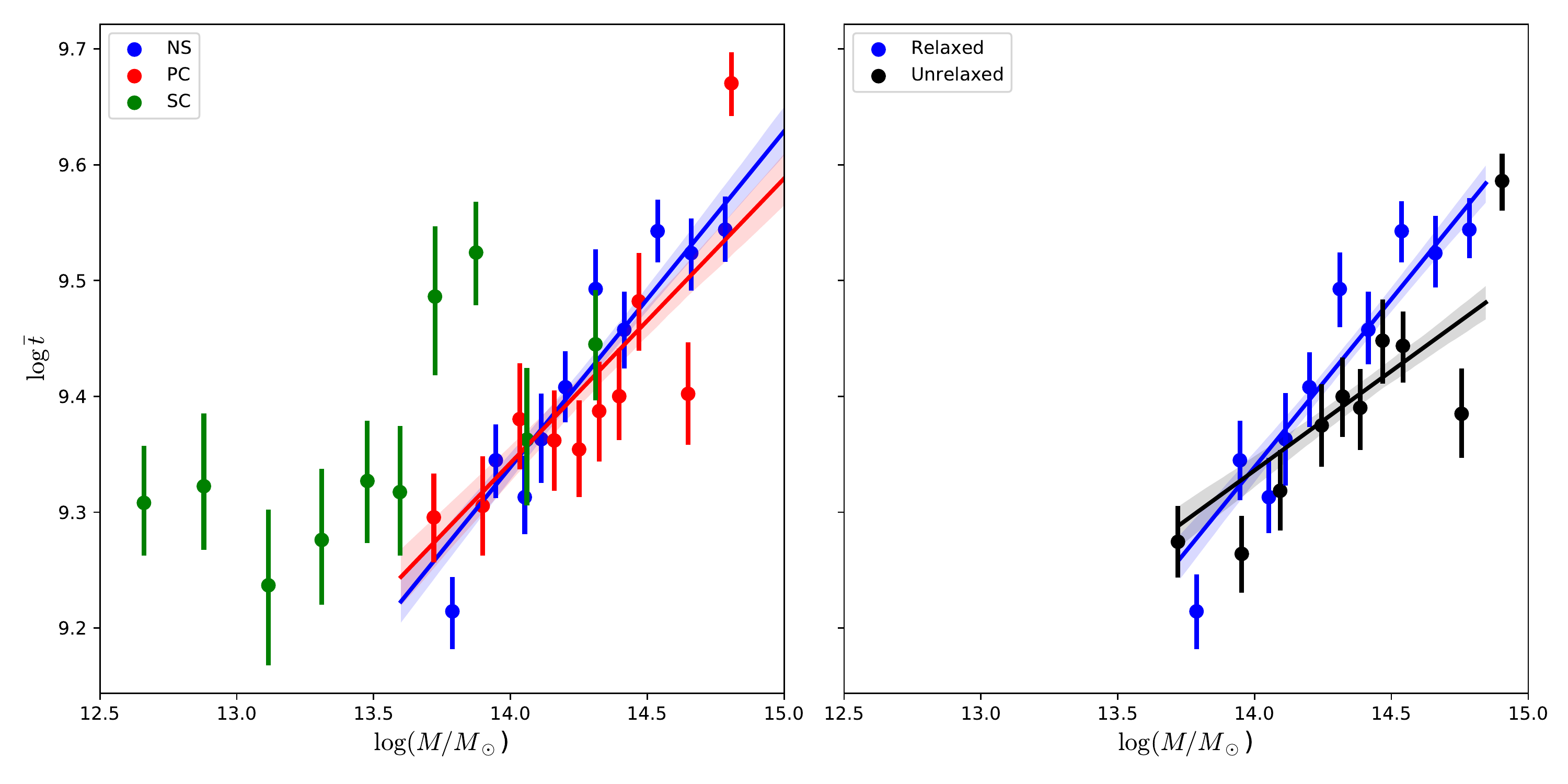}
\caption{Logarithm of mean stellar age as a function of the halo velocity dispersion (top) and halo mass (bottom). Symbols in the left plots are the same as in Figure~\ref{clusrad}. In the right plots, relaxed clusters (NS) are
represented by blue dots, and black dots indicate unrelaxed (U) clusters.}
\label{clusmhalo}
\end{figure*}

The results above suggest that the main difference between the stellar populations in individual structures comes mainly from the differences in total halo mass. In order to further explore this possibility, we plot in Figure~\ref{clusmhalo} the mean stellar age of galaxies in each structure class as a function both of the mass and the velocity dispersion of the halo they reside. As in Figure~\ref{agehis}, the right column of Figure~\ref{clusmhalo} shows the combination of PC and SC. Before separation between primary and secondary, we see a consistent age difference: Unrelaxed clusters are younger than their relaxed counterparts, even though they share the same mass and velocity dispersion range. When separating the secondaries, the age-mass and age-velocity dispersion relations become almost indistinguishable for large mass structures. This suggests that PC and NS are comparable structures, and that the main differences between clusters in different evolutionary stages (at fixed mass) come from the presence of secondaries.

\section{Discussion}

Using {\sc{LocKE}}, we have detected substructures in around 45\% of the clusters in our sample. Interpreting
the presence of substructures as a proxy for the dynamical stage of the clusters, this frequency indicates
a large proportion of unrelaxed clusters in our sample. Furthermore, the {\sc{LocKE}} performance testing
implies a family of unrelaxed clusters which our methodology is unable to identify as such. This is due
mainly to a combination of alignment of structures in the line of sight. We therefore consider the above
figure as a lower limit to the number of unrelaxed clusters in our sample. Contamination from these objects
will tend to reduce the differences between the clusters detected as relaxed and unrelaxed.

Estimating the velocity dispersion, masses and virial radius of each structure, we found that relaxed clusters
and the primary structures (i.e. the most massive) of unrelaxed clusters share approximately the same parameter distributions.
Therefore, for every unrelaxed cluster in the sample, there exists a dominant mass structure which is
comparable in mass and size to a relaxed cluster in the sample. Secondaries on the other hand are consistently
less massive and smaller. This means that the main source of unrelaxed clusters in our methodology is
the occurrence of low-mass, small systems around massive structures.

The analysis of the mean stellar ages obtained by a stellar population synthesis using STARLIGHT resulted in different age distributions for relaxed and unrelaxed clusters, with a larger contribution of younger
stellar populations in unrelaxed clusters with relation to relaxed ones. This result is in accordance with the findings of \citet{ribeiro+13a}, though these authors have used a one-dimensional test to evaluate
the dynamical stage of the clusters in their sample. Our findings are also in accordance to \citet{lopes2014}, which have found a larger proportion of blue galaxies in the outskirts of unrelaxed
clusters when compared to relaxed ones.

We then obtained the mean age distributions separating unrelaxed clusters into primaries and secondaries.
The resulting mean age difference relative to relaxed clusters are more evident for secondaries, while
primaries are much more similar to relaxed clusters. Note that we are not suggesting, at this stage, that
the status as ``secondary'' has any physical implication except the fact that it is probably contained in 
the sphere of influence of a larger cluster system. This being the case, the evolution of the whole system
will proceed and eventually reach a new equilibrium state. We have found that the main mass component
of a relaxed cluster presents a distribution of stellar populations which is similar to that of relaxed
clusters in the same mass ranges. This is not true of secondaries, and this difference may be derived
from the physical conditions the status as ``secondary'' induces in the structure or simply by the fact that the other global parameters are diverse (e.g. mass and radius).

The differences we have found in the age distributions of relaxed and unrelaxed clusters are mainly associated
to low-luminosity galaxies. Galaxies above $\sim 10^{10.8}$\,$M_\odot$ do not show any detectable sensitivity to the environment in what regards the mean stellar ages of the galaxies therein. This result is also in accordance with the
literature. A possible explanation for this effect is a scenario where the environment,
during cluster virialization, quenches the star formation in low-mass galaxies. Indeed, \citet[]{wu2014} have found that in dynamically young clusters the star formation rate is higher than in evolved clusters.
Furthermore, \citet{peng2010} have found that the stellar mass range $\log M/M\odot\lesssim 10.5$\,dex correspond to the range where the environment is the main driver of quenching. Therefore, unrelaxed clusters present lower mean stellar ages because they are richer in young low-luminosity objects, which have still not undergone the stellar population evolution characteristic of evolved systems. Because the main age differences are found for secondaries, this suggests that these peripheral, low-mass systems bring galaxies which are not representative of the environment of the primary, massive components.

Once there exists an age distribution which is specific for secondaries, how do these objects
distribute in the body of the structure they reside? We have determined how the mean stellar ages vary
as a function of the center of the structure for the three structure classes. We have found an age
gradient well established for all classes, but less intense in the secondaries. At
large radii, the mean ages are similarly low between all classes. In the central parts, however,
secondaries are consistently younger. We therefore conclude that galaxies in secondaries
are already partially pre-processed. The existence of well-defined age-clustercentric distance relations
for both relaxed and unrelaxed clusters is not new \citep[e.g.][]{ribeiro+13a}, but we
here show that the pre-processing in secondaries is not as advanced as in the primaries or in relaxed clusters. A comparable result has been found by \citet[]{Olave-Rojas18} for two massive
clusters. These authors have found that the fraction of red galaxies is higher in central regions of both clusters and substructures therein, and decrease with clustercentric distance. The authors also measured the environmental quenching efficiency, and determined that this efficiency is higher in the central regions of all structures, but particulary in the cluster inner regions. This difference may be due to the environmental processes that depend on the high density found in the central region of massive clusters, like ram-pressure or strangulation.
The difference we have found in mean ages for primaries and secondaries, in the central regions, is low ($\sim$ 1 Gyr), so that any evolutionary effects with a comparable timescale ($\sim $1-2 Gyrs) can erase this 
difference.

The findings above suggest that the mass of the structure is the major parameter behind
the mean stellar ages of its galaxy populations. We have tested this hypothesis verifying how
the mean stellar age depends on the velocity dispersion and the mass of the structure, both before
and after the decomposition of unrelaxed clusters into primaries and secondaries. We have found
age-velocity dispersion and age-mass relations well established for relaxed and unrelaxed clusters,
but these relations are distinct for these two classes. By and large, unrelaxed clusters present lower mean ages than
relaxed clusters in the same range of mass and velocity dispersion (with the possible exception of the low-
mass limit). This is in contrast with the findings of \citet{lopes2014}, which have found no
major impact of the parent halo mass on the distribution of the galaxy colors.
The difference we have found is more evident in velocity dispersion, which is more
easily contaminated by the presence of substructures, but is visible also in terms of the system
mass. When we separate primaries and secondaries, we see that primaries present relations almost
indistinguishable from relaxed clusters. Secondaries in general present mean ages close to the lower
limit for relaxed clusters, with some contamination of very old structures in the high-mass
regime, probably accounted for by an incorrect separation of components by {\sc{LocKE}}. This
reinforces the idea that primaries and relaxed clusters are similar and comparable structures,
obeying the same kinematic scaling relations. We have shown that
the secondary population is partially pre-processed, in accordance e.g. to \citet{decarvalho+17}, but this
evolution is incomplete, producing most of the observed differences in mean stellar age distribution
and radial clustercentric gradient.
These findings suggests that, as the unrelaxed clusters evolve, galaxies being accreted in the form of the secondary population will evolve due to specific mechanisms operating in the high-density of the primaries.

The exact mechanisms by which this evolution occur are uncertain.
By means of N-body plus hidrodynamical simulations of a merge between a group-size structure
and a cluster of galaxies, \citet{vija+13} have shown that both pre-processing
(due to enhanced galaxy-galaxy merger rates, ram pressure stripping in the group
environment, and tidal stripping) and post-processing (due to enhanced galaxy-galaxy collisions
and/or tidal stripping driven by the density enhancement near the pericentric passage as well
as an increased ram pressure along the shock front)
can be important for producing the stellar population distribution of cluster galaxies after
the cluster virialization. This combination of pre- and post-processing has also been invoked
by e.g. \citet{taranu+14}. These authors have found that a combination of quenching in lower mass
groups plus quenching in the cluster environment best describes
the observed colors and absorption indices of cluster galaxies when combined with a library of
subhalo orbits from N-body cosmological simulations, favouring quenching
processes with long (such as strangulation) over short timescale processes (such as ram pressure stripping). 
These works agree qualitatively with our findings in the sense that radial age gradients
are ubiquitous in our sample of low-mass secondaries, pointing to some degree of pre-processing,
but both the slope of this radial gradient and the mean stellar age of galaxies therein are
lower than in relaxed clusters, indicating that further evolution has to operate after
the parent haloes merge. On the other hand, \citet{delucia+12}, comparing galaxy merger trees with
optical colors of group and cluster galaxies,
have found that more massive haloes present a higher fraction of quenched galaxies due to
the fact that satellites of massive haloes have been satellites -- and therefore subject
to environmental quenching -- for a longer time than
satellites of low-mass haloes. As a consequence, the fraction of quenched galaxies must scale with
the parent halo mass. This helps to explain the correlation we have found between the structure
mass and mean stellar age for relaxed clusters and primaries. However, this would imply that the
observed differences between the stellar populations of secondary structures and relaxed clusters is not due to an increased quenching efficiency in high mass systems -- i.e. quenching could proceed
for satellite galaxies at the same rate as before the merging.
Both scenarios -- including or not an increased post-processing efficiency -- could
be reconciled with our results  due to the small differences we have found in mean stellar ages, as long as the scaling relations after unrelaxed cluster virialization
evolve quickly enough to match those of relaxed clusters. A direct test of the occurrence
of increased post-processing in our sample can be made by investigating the properties of galaxies
in unrelaxed clusters as a function of their projected position with relation to the putative
shock front and the evolutive stage of the merge. This analysis will be presented in a forthcoming paper.

\section{Conclusions}

We have investigated the occurrence of unrelaxed systems in a sample of 408 clusters of galaxies drawn
from the \citet[]{tempel2012} catalogue. For substructure detection, extraction and automatic estimation of the kinematic parameters we devised the code LocKE (Local Kinematic Estimator) and tested its performance against another
multidimensional decomposition code used to detect substructures in clusters of galaxies (mclust). The
individual structures in unrelaxed clusters have been separated into primaries and secondaries
according to their dynamical mass. The mean stellar ages of galaxies in relaxed and unrelaxed clusters have been derived using the results of a stellar population synthesis with the STARLIGHT code. Our main findings can
be summarized as follows:

\begin{itemize}
\item We found a frequency of $\sim$ 45\% of substructures in the sample clusters, which is in agreement with the literature. Given the intrinsic incompleteness of LocKE for structures aligned with the line of sight,
this figure must be seen as as upper limit to the real frequency of unrelaxed clusters in the sample.
\item Primary (i.e. the most massive) structures in unrelaxed clusters present a similar mass distribution as
relaxed clusters, while secondaries are much smaller and less massive.
\item In unrelaxed clusters, the galaxies present a lower mean stellar age than galaxies in relaxed clusters. When separating unrelaxed clusters into primaries and secondaries, we find that most of this difference is due
to lower mean stellar ages in secondaries.
\item The age differences above are detected for all galaxy stellar masses, but seem to be less marked
at the highest stellar mass bins ($\gtrsim 10.8$\,dex).
\item A mean age-radius relation is observed both for relaxed clusters and for primaries and secondaries
in unrelaxed clusters. The slope of the relation is, however, less steep for secondaries. This suggests that
substructures in clusters bring to the system galaxies which are already pre-processed at the group
scale, but this pre-processing is still incomplete and will proceed after the galaxies penetrate into
the dense regions of the primaries.
\item Relaxed and unrelaxed clusters describe different mean age - mass and mean age - velocity dispersion
relations. However, the same relations are nearly identical for primaries and relaxed clusters. We interpret
our findings as an  evidence that the large-scale merging of galaxy systems does not have a dramatic impact on the stellar populations of galaxies in primary substructures. On the contrary, the observed trends can be explained
as a result of the infall, onto large systems, of low-mass systems which are, due to its reduced mass,
richer in galaxies with younger stellar populations. The virialization process of unrelaxed clusters
is therefore accompanied by the evolution of the stellar populations of galaxies in these infalling groups.

\end{itemize}

\section{Acknowledgements}

We thank FAPERGS and CNPq for financial support.
This study was financed in part by the Coordena\c{c}\~ao de Aperfei\c{c}oamento de Pessoal de N\'ivel Superior -- Brasil (CAPES) -- Finance Code 001.

Funding for SDSS-III has been provided by the Alfred P. Sloan Foundation, the Participating Institutions, the National Science Foundation, and the U.S. Department of Energy Office of Science. The SDSS-III web site is http://www.sdss3.org/.

SDSS-III is managed by the Astrophysical Research Consortium for the Participating Institutions of the SDSS-III Collaboration including the University of Arizona, the Brazilian Participation Group, Brookhaven National Laboratory, Carnegie Mellon University, University of Florida, the French Participation Group, the German Participation Group, Harvard University, the Instituto de Astrofisica de Canarias, the Michigan State/Notre Dame/JINA Participation Group, Johns Hopkins University, Lawrence Berkeley National Laboratory, Max Planck Institute for Astrophysics, Max Planck Institute for Extraterrestrial Physics, New Mexico State University, New York University, Ohio State University, Pennsylvania State University, University of Portsmouth, Princeton University, the Spanish Participation Group, University of Tokyo, University of Utah, Vanderbilt University, University of Virginia, University of Washington, and Yale University. 











\bsp	
\label{lastpage}
\end{document}